\documentclass{PoS}
\usepackage{url}
\usepackage[utf8]{inputenc}

\title{Hints and challenges in heavy flavor physics}

\ShortTitle{Hints and challenges in heavy flavor physics}

\author{\speaker{Shoji Hashimoto}\\
  High Energy Accelerator Research Organization (KEK), 
  Tsukuba 305-0801, Japan
  \\
  SOKENDAI (Graduate University for Advanced Studies),
  Tsukuba 305-0801, Japan
  \\
  E-mail: \email{shoji.hashimoto@kek.jp}}

\abstract{Heavy flavor physics entered a new era when the Belle II
  experiment observed its first collision. 
  There are several hints found so far by BaBar, Belle, and LHCb in
  particular, that suggest the physics beyond the Standard Model
  appearing in the loop processes at short distances.
  They will be further tested by higher precision experiments in the
  coming years, while the role of lattice QCD is to understand the
  long-distance physics quantitatively so that one can unambiguously
  isolate the short-distance physics from the experimental data. 
  I summarize the status towards this goal and then look at the
  challenges we are facing.}

\FullConference{The 36th Annual International Symposium on Lattice Field Theory - LATTICE2018\\
                22-28 July, 2018\\
                Michigan State University, East Lansing, Michigan, USA.}

\begin{document}

\section{Introduction}
\label{sec:Introduction}
On April 26, 2018, the Belle II detector observed the first collision
of an electron-positron pair accelerated by the SuperKEKB accelerator.
This is the event that opens a new era of heavy flavor physics.
The SuperKEKB is designed to deliver 50 times more luminosity than 
the previous KEKB accelerator did.
At the time of the conference, the Phase 2 operation of SuperKEKB,
which is the run with Belle II but without its inner-most vertex
detectors, 
had just been finished and the machine is scheduled to restart in
March 2019 with fully equipped Belle II.
The plan then is to accumulate 50~ab$^{-1}$ within 5--6 years of
running. 

In $B$ meson decays, there are already several hints of new physics.
A well-known example is the $B\to D^{(*)}\tau\nu$ decay branching
fractions, which show an enhancement compared to the Standard Model
expectation \cite{Amhis:2016xyh,HFLAV_BtoDtaunu}.
Because of $\tau$ lepton mass, the form factor uncertainty does not
completely cancel even in the ratio between the branching fractions
for $B\to D^{(*)}\tau\nu$ and $B\to D^{(*)}\mu\nu$, and the lattice
calculations by 
Fermilab/MILC \cite{Lattice:2015rga} and 
HPQCD \cite{Na:2015kha}
have been used to predict the Standard Model value.
The experimental data show a tension of about 3.8$\sigma$.

For the rare decays $B\to K^{(*)}\ell^+\ell^-$, some hints of new
physics have been found in the lepton flavor universality test,
$R(K^{(*)})=
\Gamma(\bar{B}\to\bar{K}^{(*)}\mu^+\mu^-)/
\Gamma(\bar{B}\to\bar{K}^{(*)}e^+e^-)$
(the most precise LHCb observations to date are 
\cite{Aaij:2014ora,Aaij:2017vbb};
see also the references therein)  
as well as an angular asymmetry $P_5'$ of $B\to K^*\mu^+\mu^-$
(results from LHCb \cite{Aaij:2015oid} and 
Belle \cite{Wehle:2016yoi}, 
CMS \cite{Sirunyan:2017dhj} and 
ATLAS \cite{Aaboud:2018krd}).

A more relevant quantity to the lattice calculation
is the branching fraction of the $B\to K\ell^+\ell^-$ decay mode.
The differential decay rate is estimated using a lattice calculation
and other constraints.
The experimental data by CDF, Belle, BaBar and LHCb look consistently
lower as shown in Figure~6 of \cite{Du:2015tda}. 
This comparison is made away from the region of charmonium resonances
($J/\psi$ and $\psi'$), since there are huge contributions from 
$B\to \psi^{(')} K\to \ell^+\ell^- K$ due to a Cabbibo allowed process 
$b\to c\bar{c}s$.
Quantitative estimate of such long-distance contribution is still
missing, and a separation of short-distance physics (new physics) and
long-distance physics (QCD) remains a difficult problem.

The challenges for lattice QCD in the context of heavy flavor physics
are twofold.
One is the precision frontier.
For the simple quantities like decay constants and semileptonic form
factors, one wants to achieve the precision as good as that the
corresponding experiments provide.
The other is the study to understand unknowns, which includes the
long-distance effect as well as the puzzles between the inclusive and
exclusive determination of $|V_{cb}|$ and $|V_{ub}|$.
For such quantities, the way to proceed might not be unique, and even
a crude calculation could be helpful.

In this talk I describe my (sketchy) understanding of the status of these
fronts. 
Section~\ref{sec:precision} summarizes the status of the precision
frontier.
I take the most recent calculation of the $B$ meson decay constant as
an example
in order to consider about what has been and what will be the main
challenges.
I then move to the discussion of the exclusive determination of
$|V_{cb}|$ in Section~\ref{sec:Vcb}.
I focus on this quantity because there was a significant progress and
associated confusions in the last couple of years.
Section~\ref{sec:challenges} is devoted to more complicated
quantities, such as the inclusive $B$ meson decay calculation and the
long-distance effect in $B\to K\ell^+\ell^-$.

I have to mention that this is not a comprehensive review.
I am not going to provide averages of lattice calculations.
For the averages, the Flavour Lattice Averaging Group (FLAG) reviews 
most recent calculations and provide up-to-date averages every 2--3
years.
The last edition (FLAG3) was published in 2016 \cite{Aoki:2016frl} and
the next edition (FLAG4) is scheduled to appear in early 2019
\cite{Aoki:2019cca}.

\section{Precision frontier}
\label{sec:precision}
The biggest news in the precision frontier in the last couple of years
came from the Fermilab/MILC collaboration, which reported a
calculation of the $B$ meson decay constant at the level of precision
better than 1\% for the first time \cite{Bazavov:2017lyh}.
The result is $f_B$ = 189.4 $\pm$ 1.4~MeV, which shows a significant
improvement compared to the previous average by FLAG
\cite{Aoki:2016frl},
192.0 $\pm$ 4.3~MeV for $N_f=2+1$ or
186 $\pm$ 4~MeV for $N_f=2+1+1$
(based on the calculations of 
Fermilab/MILC \cite{Bazavov:2011aa},
RBC \cite{Christ:2014uea,Aoki:2014nga},
and HPQCD \cite{Na:2012kp,Dowdall:2013tga}).

\subsection{Discretization effect for heavy}
Achieving such precision is highly non-trivial, because the
heavy quarks are much harder to treat on the lattice due to their
short Compton wavelength.
This is indeed the reason of long history of works to develop and test 
effective theories for heavy quarks on the lattice.
The solution adopted in this work, on the other hand, is to use the
Highly Improved Staggered Quark (HISQ) action \cite{Follana:2006rc}
for both heavy and light quarks, and to extend the simulations on the
lattices with small lattice spacings down to 
$a$ = 0.042~fm at the physical pion mass.
The computational effort spent for this work,
70~TFlops$\times$yr for the ensemble generation plus another
70~TFlops$\times$yr for the measurements (both in the unit of
sustained TFlops),
was of course a key for the success, 
but does not explain everything because the highest lattice cutoff is
still around 5~GeV and not enough to satisfy the condition 
$am_b\ll 1$ for the physical $b$ quark mass $m_b$.

The discretization effect for heavy quark may be estimated according
to the so-called Fermilab interpretation
\cite{ElKhadra:1996mp}, that applies the Heavy Quark Effective Theory
(HQET) for the lattice regularization with finite $am_b$.
The leading discretization effect for heavy HISQ quarks is then
estimated to be of $O(\alpha_sa^2m^2)$.
Numerically, it is not small for the physical $b$ quark mass.
Even for a fictitious ``$b$'' quark that has an intermediate mass of
3~GeV, it gives a rough estimate for the discretization effect to be
about 7\%. 
(Other parameters are taken as
$\alpha_s\sim 0.2$ and $a^{-1}\simeq$ 5~GeV, which
correspond to the Fermilab/MILC calculation.)

In order to achieve the sub-\% precision, therefore, the
discretization error has to be eliminated by a continuum
extrapolation. 
In the calculation of the Fermilab/MILC collaboration, this is done
with a number of lattice data calculated at various ``$b$''
quark masses between the charm quark mass and the physical $b$ quark
mass.
The lattice spacing covers a wide range from 0.15~fm down to 0.03~fm,
which enable them to fit the data globally allowing the terms of 
$1/m^k$ ($k\geq 0$) and $\alpha_s(ma)^n$ ($n\geq 2$) to account for
the physical $1/m$ dependence as well as the discretization effects.
The heavy quark mass one can reach at each lattice spacing without too
large discretization effect and can be used in the analysis is
limited by a condition that $ma$ is below some value.
The Fermilab/MILC collaboration chose $ma<0.9$.

\begin{figure}[tbp]
  \centering
  \includegraphics[width=10cm]{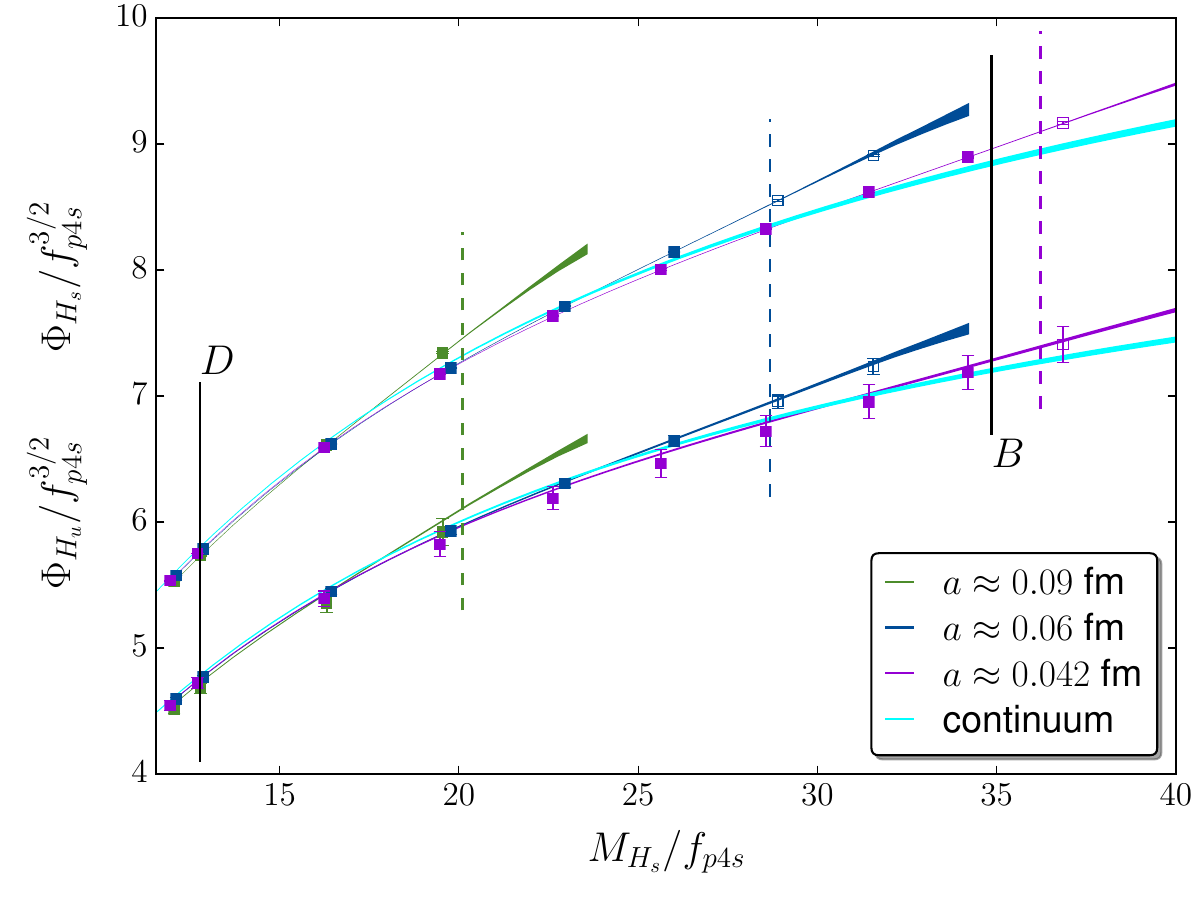}
  \caption{Heavy quark and continuum extrapolation of heavy meson
    decay constants $\Phi_H\equiv f_H\sqrt{m_H}$.
    The plot is from Fermilab/MILC \cite{Bazavov:2017lyh}.
    Two bunches of curves represent those for $B_s$ (above) and 
    $B_u$ (below). 
    For each case, the results at various heavy quark masses obtained
    at three different lattice spacings are plotted.
    The continuum limit obtained by a global fit is shown by cyan
    curves. 
  }
  \label{fig:fB}
\end{figure}

Fig.~\ref{fig:fB} demonstrates this global fit.
One can see that the results at growing lattice cutoffs tend
to follow a single curve more closely, and they finally show the
envelope to represent the continuum limit.
The cyan curve is their estimate of the continuum limit obtained by
the global fit.
A surprising observation is that the deviation from this continuum
limit is much smaller than what we estimate using the effective
theory.
For instance, for our fictitious $b$ quark at ``$m_b$'' = 3~GeV, the
meson mass is roughly in the middle between the physics $M_D$ and
$M_B$.
The deviation from the continuum limit (cyan) seems to be much smaller
than the estimated 7\% at $a\simeq$ 0.042~fm (blue) and is
almost invisible in the scale of Fig.~\ref{fig:fB}. 
(The size of the discretization effect is actually about 3\% or less
even including the coarser lattices of $a\simeq$ 0.09~fm.)

This small discretization effect may be partly due to the fact that
the renormalization constant of the axial-vector current is
automatically determined when one uses the same relativistic quark
action for both heavy and light.
The bulk of the discretization effect of $O(\alpha_s a^2m^2)$ may be
absorbed in this method.
This might be a {\it lucky} situation that happens only for the decay
constant and might not be the case for other quantities.

I also mention that this strategy to employ the HISQ heavy quark even
for ``bottom'' and to extrapolate to the physical point originated
from the pioneering work by the HPQCD collaboration
\cite{McNeile:2011ng}. 
The Fermilab/MILC collaboration extended this program to smaller
lattice spacings and to higher statistics, and finally achieved the
precision calculation.

There are various strategies for the treatment of heavy quarks on the
lattice. 
For the effective theory approaches, 
such as the NRQCD action \cite{Lepage:1992tx}, the
Fermilab action \cite{ElKhadra:1996mp} and its variants, 
one has to match the action and operators to reproduce the
relativistic continuum fermion.
The matching is usually carried out using perturbation theory, and the
error of $O(\alpha_s^2)$ remains when the one-loop corrections are
included. 
(Note also that this can be improved by the so-called mostly
non-perturbative matching, which means that the currents made of
degenerate quarks are non-perturbatively renormalized and possible
deviations for non-degenerate quarks are estimated by perturbation
theory.) 
The method to perform the matching completely non-perturbatively has
been formulated by the ALPHA collaboration \cite{Heitger:2003nj,Blossier:2010jk,Blossier:2010vz,Blossier:2010mk}.
It requires dedicated simulations with various volumes and lattice
spacings, and takes time to carry out.
Finally, the relativistic approach has become more common as the
computational power increases.
While one needs small lattice spacings in order to suppress
discretization effects, 
the non-perturbative matching of the operators is much simpler than
the effective theory approaches.
Such a brute-force approach could be a simple solution in some cases,
and the most recent Fermilab/MILC calculation of $f_B$ is a prominent
example.

\subsection{Growing noise and excited-state contamination}
\begin{figure}[tbp]
  \centering
  \vspace*{-15mm}
  \includegraphics[width=7.5cm]{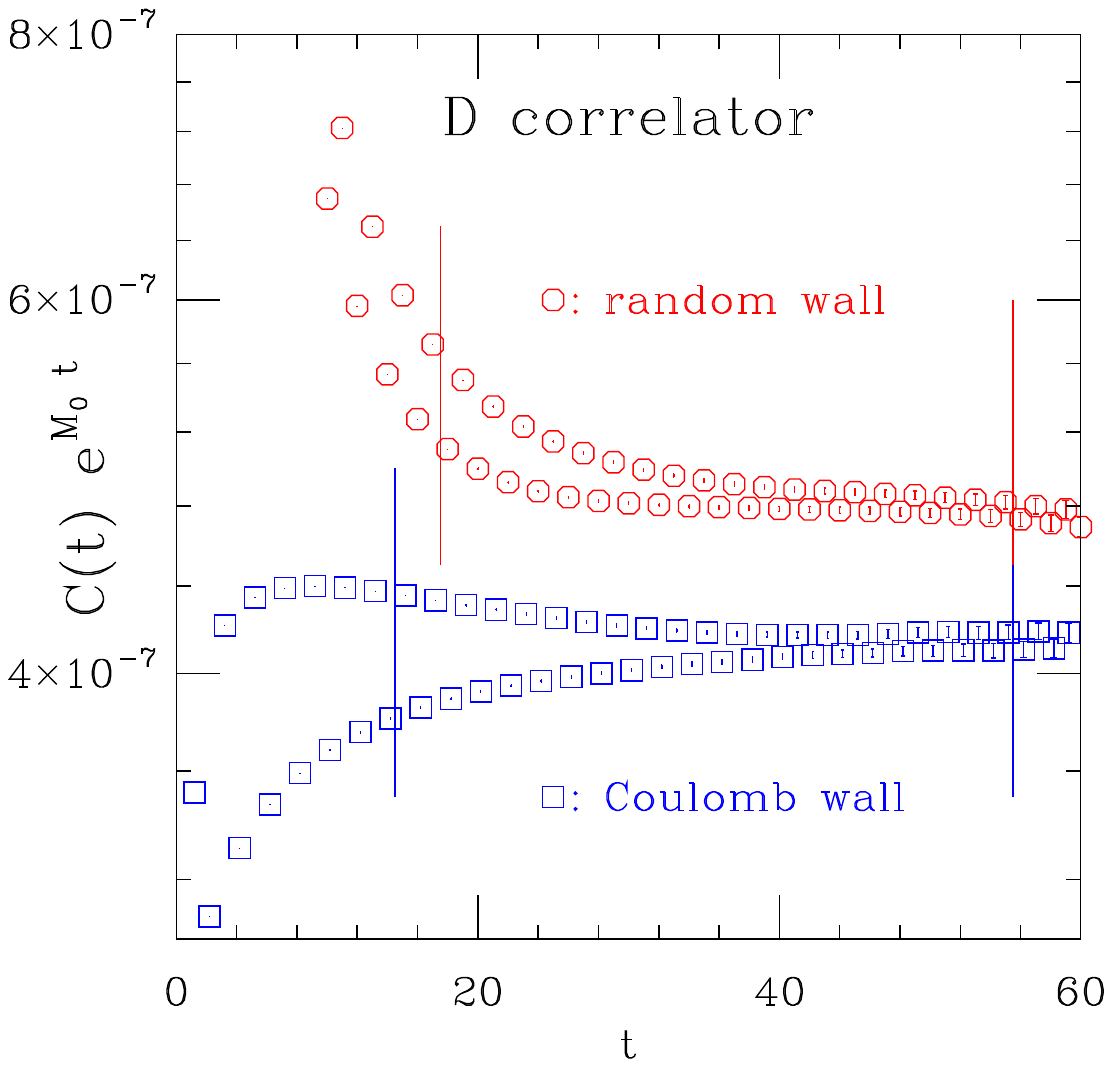}
  \includegraphics[width=7.5cm]{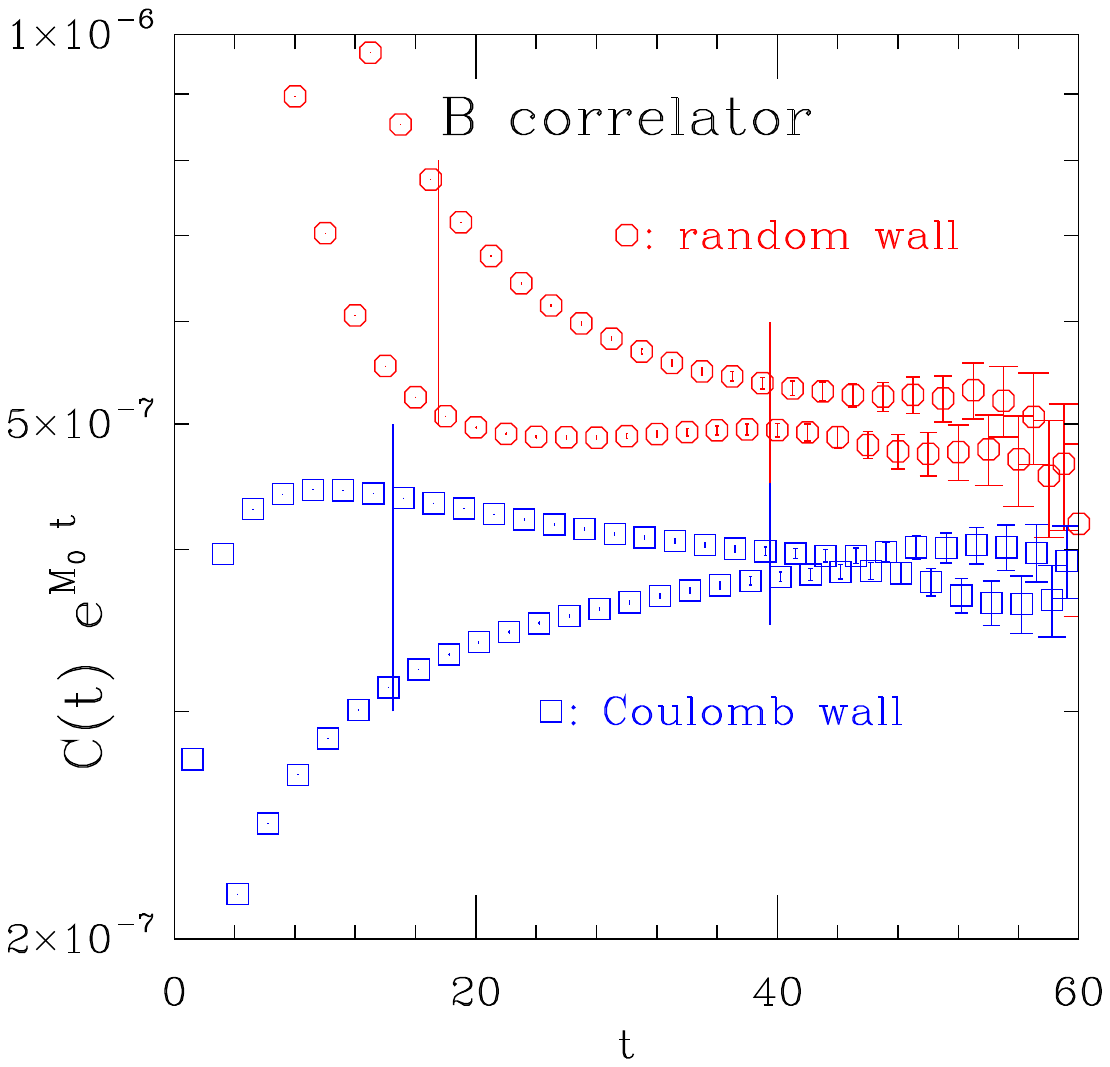}
  \vspace*{-20mm}
  \caption{
    $D$ meson (left) and $B$ meson (right) correlator divided by
    their ground-state exponential function $e^{-M_0t}$.
    The plots are from Fermilab/MILC \cite{Bazavov:2017lyh}.
    They are obtained on a fine ($a\simeq$ 0.042~fm) lattice and at
    physical pion mass.
    Circles and squares are with different sources, and the zigzag
    structure is due to the staggered fermion.
  }
  \label{fig:DandBcorr}
\end{figure}

The statistical error poses more problems for heavy quarks.
According to Lepage's analysis \cite{Lepage:1989hd}, 
which was numerically confirmed in \cite{Hashimoto:1994nd},
one can predict how rapidly the statistical noise grows in the
correlation functions calculated on the lattice.
For the heavy-light meson correlator $C_{HL}(t)$, the statistical
noise $\delta C_{HL}(t)$ grows as
\begin{equation}
  \frac{\delta C_{HL}(t)}{C_{HL}(t)} \propto
  \exp\left[
    \left( m_B - \frac{m_{\eta_b}+m_\pi}{2} \right) t
  \right].
\end{equation}
The noise ``doubling time''\footnote{Actually the time that the noise
  becomes $\times$2.7.}
$(m_B-(m_{\eta_b}+m_\pi)/2)^{-1}$
is determined by the physical mass spectrum and there is
nothing to do with the lattice details.
Numerically, it is about 0.4~fm for the $B$ meson, which is much worse
compared to the $D$ meson (0.65~fm).
The growing noise, especially for the $B$ meson, can be clearly seen in
the data by Fermilab/MILC \cite{Bazavov:2017lyh}, which is reproduced
in Figure~\ref{fig:DandBcorr}.

As a result, one is forced to fit the correlator before the ground
state dominates, in order not to lose the signal at all.
Such a fit including the excited-state contamination modeled by
multiple exponential functions is a numerically difficult problem.
In the $f_B$ calculation by Fermilab/MILC \cite{Bazavov:2017lyh}, 
the multi-exponential fit is applied with five distinct states
included. 
The fit range is shown by vertical lines in the plots; 
the maximal time separation fitted is about 1.6~fm for the $B$ meson
(right) where the plateau is not yet reached.
Two of the states out of five included in the fit are opposite-parity
states due to the use of the staggered fermion, 
and the other two are the excited states of the target $B$ meson.
Assuming the physical mass spectrum experimentally observed, 
the candidates for these low-lying excited states are the two-body
$B^*\pi$ states with finite relative momenta between $B^*$ and $\pi$.
On a large lattice of size $L\simeq$ 6~fm, 
as used by the Fermilab/MILC calculation,
they start from 300~MeV above the ground state and their spectrum is
nearly continuum like, {\it i.e.}
a dense distribution of different energy eigenstates are expected.
The excited-state energy obtained by the fit is consistent with this
band of states, but they are not separately identified.
The extraction of the ground state would be stable even though the
excited states are not precisely accounted for, but the crucial
question is whether the statistical error thus estimated is valid
especially when the plateau is not cleanly identifiable.
Such careful study would become more relevant for many quantities as
the numerical precision improves.

The lattice simulation at the physical pion mass is a big challenge
for us if not a nightmare. 
First of all, with the small pion mass the signal-to-noise ratio
becomes worse for the reason just described, 
and one has to fit the lattice correlators before plateau is reached. 
This itself is fine, but the situation gets even worse because the
excited-state energy is made lower as $m_\pi$ and the isolation of the
the ground state needs larger time separations.
These excited states become dense when the lattice volume is large as
required to keep the finite volume effect under control.
It means that the multi-exponential function is no longer a valid
description of the correlator.

This problem had already become manifest for baryons.
The analysis of the noise growth rate implies that the 
``doubling time'' is about 0.27~fm for nucleon, even smaller than for
the $B$ meson.
In a recent computation of nucleon correlator at the physical point, 
{\it i.e.} the one by the PNDME collaboration \cite{Rajan:2017lxk}, 
no clear plateau is visible before signal dies at around the time
separation of 1.5~fm.
The authors attempted multi-exponential fits, and 
the resulting excited-state energy was not consistent with what one
expects from the $N\pi$ continuum states.
An interpretation might be that the interpolating operator used in
this work has small overlap with such two-body states, and even
higher-energy states show up in the data.
But, then, one should ask if there is any bias due to the unseen
continuum-like states.
Certainly, a better understanding of the correlator at short distances 
is desired.

Another important warning may be found in the computation of the axial
charge of nucleon.
B\"ar gave a plenary talk at Lattice 2017 \cite{Bar:2017gqh}
concerning a possible bias due to the correlator fit at too short
source-sink separations
(see also \cite{Bar:2016uoj,Bar:2018wco} 
and a talk at this conference \cite{Bar:2018akl}).
Within chiral perturbation theory (ChPT) one can compute the
contribution of the $N\pi$ states to the correlator relevant to the
calculation of the nucleon axial charge $g_A$.
With finite source-sink separations, but larger than 2~fm, he
estimated a bias that overestimates $g_A$ by about 2--7\%.
Many of the lattice computations, on the other hand, gave results 
{\it lower} than the experimental value by about 3--10\%.
A potentially serious problem here is that the source-sink separation
adopted in these calculation is too small, $\lesssim$ 1.5~fm, to
safely apply ChPT.
(The separation between the source (or sink) and the current is less
than half of this.)
Since there is no a priori criterion for the necessary time
separation, 
the nucleon $g_A$ would provide an important benchmark problem for
lattice computation, for which the consistency with the theoretical
estimate has not been established.

A similar, or even tougher, problem is the calculation of heavy baryon
decay form factors, such as those of $\Lambda_b\to p\ell\nu$.
The noise doubling time for $\Lambda_b$ is about 0.25~fm, which is
even shorter than that for nucleon.
Because of such a bad signal-to-noise, the authors of
\cite{Detmold:2015aaa}
had to fit the correlators at the source-sink separations between 0.5 
and 1~fm, much smaller than the separation typically adopted in the
$B\to\pi\ell\nu$ form factor calculations.
They numerically confirmed the stability of the ground-state signal
within the existing data, but more stringent consistency checks
including those of excited-state energy would be highly desired since
the fit range does not even touch the plateau region.

\subsection{Heavy-to-light decays}
Semileptonic decays of heavy mesons to light mesons, such as
$B\to\pi\ell\nu$ and $B_s\to K\ell\nu$,
are challenging quantities for the lattice calculation.
It is not only due to the heavy initial $B$ meson but the final
state ($\pi$ or $K$ for these examples)
can have relatively large momenta.
With a finite momentum, the noise growth rate of pion correlator, for
instance, is given as
$\delta C_\pi(t,\mathbf{p})/C_\pi(t,\mathbf{p})\propto
\exp[(E_\pi(\mathbf{p})-m_\pi)t]$,
and it sets the limit on the value of the momentum transfer $q^2$ one
can reach.
The maximum momentum of the final state pion in the lattice
calculation is typically around
600--800~MeV$/c$, which is much lower than the kinematically allowed
maximum $\sim$ 2.6~GeV$/c$.
Previous calculations, such as those by
RBC/UKQCD \cite{Flynn:2015mha}, 
Fermilab/MILC \cite{Lattice:2015tia},
ALPHA \cite{Bahr:2017ivh}, 
actually observed the growing noise for larger momenta.
These groups therefore restricted themselves for small recoil momenta
(and thus large $q^2$) and fit the plateau.
The possibility of fitting including much shorter time-separations 
even without finding plateaus has not been fully explored. 

At this conference, there have been presentations on the
heavy-to-light form factor calculations by
Fermilab/MILC (Gelzer {\it et al.} \cite{Gelzer_lat18}, see also
\cite{Gelzer:2017edb,Lattice:2017vqf} for preliminary results at
Lattice 2017),
RBC/UKQCD (Witzel {\it et al.} \cite{Witzel_lat18}),
JLQCD (Colquhoun {\it et al.} \cite{Colquhoun:2018kwj}, see also
\cite{Colquhoun:2017gfi}),
HPQCD (Bouchard {\it et al.} \cite{Bouchard_lat18}, see also
\cite{Monahan:2018lzv} for a full paper),
as well as
ETM (Riggio {\it et al.} \cite{Lubicz:2018scy}, see also 
\cite{Lubicz:2018rfs} for a full paper).

To conclude this section, I emphasize that precise calculation has
become realistic even for $b$ quark.
An important factor for this development is the use of the HISQ action
combined with a large amount of computer resources.
Results with other lattice formulations are highly desired to
cross-check the results.
Extension of the calculation to more complicated problems, such as the
heavy-to-light form factors, is more challenging mainly due to the
noise problem.
The ``fit-before-plateau'' strategy will be used more commonly for
these quantities, and we need thorough understanding of the
excited-state contaminations especially those from the continuum, such
as  $B^{(*)}\pi$, states.

\section{A $|V_{cb}|$ story}
\label{sec:Vcb}

One of the key CKM elements, $|V_{cb}|$, can either be determined
using the exclusive decay processes $B\to D^{(*)}\ell\nu$ or
the inclusive decay rate of $B\to X_c\ell\nu$.
The exclusive determination relies on the constraints given by heavy
quark symmetry and lattice calculations.

In the limit of infinitely heavy quarks, 
$m_b$, $m_c\to\infty$,
the heavy-to-heavy meson transition form factors, 
for both $B\to D\ell\nu$ and $B\to D^*\ell\nu$,
can be written in terms of an universal function $\xi(w)$, 
called the Isgur-Wise function \cite{Isgur:1989vq,Isgur:1989ed}.
This is a remarkable result of the heavy quark symmetry.
Here, the argument $w$ is an inner product of initial and final meson 
velocities, $w\equiv v\cdot v'$.
In the zero-recoil limit, $w=1$, the process is nothing but an
insertion of a temporal vector current between static mesons, and the
form factor is normalized to one due to vector current conservation:
$\xi(1)=1$. 
The correction to the $m_{b,c}\to\infty$ limit starts from $O(1/m^2)$,
{\it i.e.} no $O(1/m)$ corrections exist \cite{Luke:1990eg},
in the zero-recoil limit, so that the size of the correction is
suppressed by a factor of $\bar\Lambda^2/m_c^2$ with $\bar\Lambda$ a
typical QCD scale and thus expected to be about 10\% or less.
The standard strategy for a precise determination of $|V_{cb}|$ is
therefore to measure $|V_{cb}|{\cal F}(w)$ in the experiments 
(${\cal F}(w)$ stands for the form factor at finite $m_{b,c}$).
One extrapolates the data to the $w\to 1$ limit and inputs an
theoretical estimate for ${\cal F}(1)$ to extract $|V_{cb}|$.
In this way one can avoid hadronic uncertainty as much as possible.

The best theoretical estimate for ${\cal F}(1)$ now comes from lattice 
calculation, and the lattice calculation itself gets benefits from the
heavy quark symmetry.
For instance, a double ratio of the zero-recoil form factors
\begin{equation}
  |h_+(1)|^2 = \frac{
    \langle D|\bar{c}\gamma_0 b|\bar{B}\rangle
    \langle\bar{B}|\bar{b}\gamma_0 c|D\rangle
  }{
    \langle D|\bar{c}\gamma_0 c|D\rangle
    \langle\bar{B}|\bar{b}\gamma_0 b|\bar{B}\rangle
  }
\end{equation}
becomes unity in the heavy quark limit.
The left-hand side doesn't have a denominator because the current
conservation of the flavor-conserving vector current normalizes 
the corresponding form factor to 1.
Away from the heavy quark limit, a symmetry under the exchange between
$b$ and $c$ forbids the correction of $O(1/m)$.
By designing the lattice calculation such that this symmetry is
manifest, one essentially calculates the deviation from the heavy
quark limit, $|h_+(1)|^2-1$, which is of $O(1/m_{b,c}^2)$
\cite{Hashimoto:1999yp,Hashimoto:2001nb}.

Unquenched lattice results for the zero-recoil form factor have so far
been obtained by 
Fermilab/MILC
(for $B\to D\ell\nu$ \cite{Lattice:2015rga} and 
for $B\to D^*\ell\nu$ \cite{Bailey:2014tva}) 
and 
HPQCD
(for $B\to D\ell\nu$ \cite{Na:2015kha} and 
for $B\to D^*\ell\nu$ \cite{Harrison:2017fmw}).
Combined with the experimental averages provided by the HFLAV group
\cite{Amhis:2016xyh},
the results for $|V_{cb}|$ are
0.0398(10)(14) from $B\to D\ell\nu$ and 0.0391(5)(5) from $B\to D^*\ell\nu$.
It is to be compared with another determination,
0.0422(8),
from the inclusive $B$ meson decays \cite{Amhis:2016xyh}.
If we take them at their face values, there is a tension of about
$3\sigma$ between the exclusive and inclusive determinations.
This is a long-standing puzzle in the determination of $|V_{cb}|$ that
prevents us from performing more precise tests of the CKM unitarity.

More recently, the situation was changed by a new analysis of 
$B\to D^*\ell\nu$ by Belle \cite{Abdesselam:2017kjf}.
They created an {\it unfolded} data of the differential decay rate
available in the form that theorists can attempt their own analysis.
Several theorists have actually studied the fit of the experimental
data of $|V_{cb}|^2{\cal F}^2(w)$ using various ansatz for the
functional form of ${\cal F}(w)$, and found that the results depend on
the details of the fit function
\cite{Bigi:2017njr,Grinstein:2017nlq,Bernlochner:2017xyx}.

The commonly used fit ansatz in the experimental analyses had been
that of Caprini, Lellouch, Neubert (CLN) \cite{Caprini:1997mu},
which was developed from a more general formulation by
Boyd, Grinstein, Lebed (BGL) \cite{Boyd:1995cf}.
The both use the idea of the dispersive bound
\cite{Okubo:1972ih,Bourrely:1980gp},
which relates two-point functions of the form
$\langle J^{\bar{c}b}J^{\bar{b}c}\rangle$
to a sum of the matrix elements
$|\langle 0|J^{\bar{c}b}|\bar{B}^{(*)}_i D^{(*)}_j\rangle|^2$.
Here, the subscripts $i$ and $j$ denote possible states allowed by
symmetries.
Since the matrix elements are written in terms of the corresponding
form factors, this relation gives certain constraints on the form
factors integrated over their kinematical variables.
The constraints are used such that the functional form of 
${\cal F}(w)$, for instance, is well approximated by a polynomial of a 
variable
$z=(\sqrt{1+w}-\sqrt{2})/(\sqrt{1+w}+\sqrt{2})$
truncated at some order.
On top of that, the CLN ansatz incorporates some estimates of the
$1/m$ corrections from heavy quark effective theory, and thus
puts tighter constraints than the BGL does.
For instance, the form of $h_{A_1}(w)$, 
a dominant part of ${\cal F}(w)$, 
is parametrized by only two parameters up to $O(z^3)$,
{\it i.e.} the slope of ${\cal F}(w)$ at $w=1$ determines its
curvature too. 

\begin{figure}[tbp]
  \centering
  \includegraphics[width=10cm]{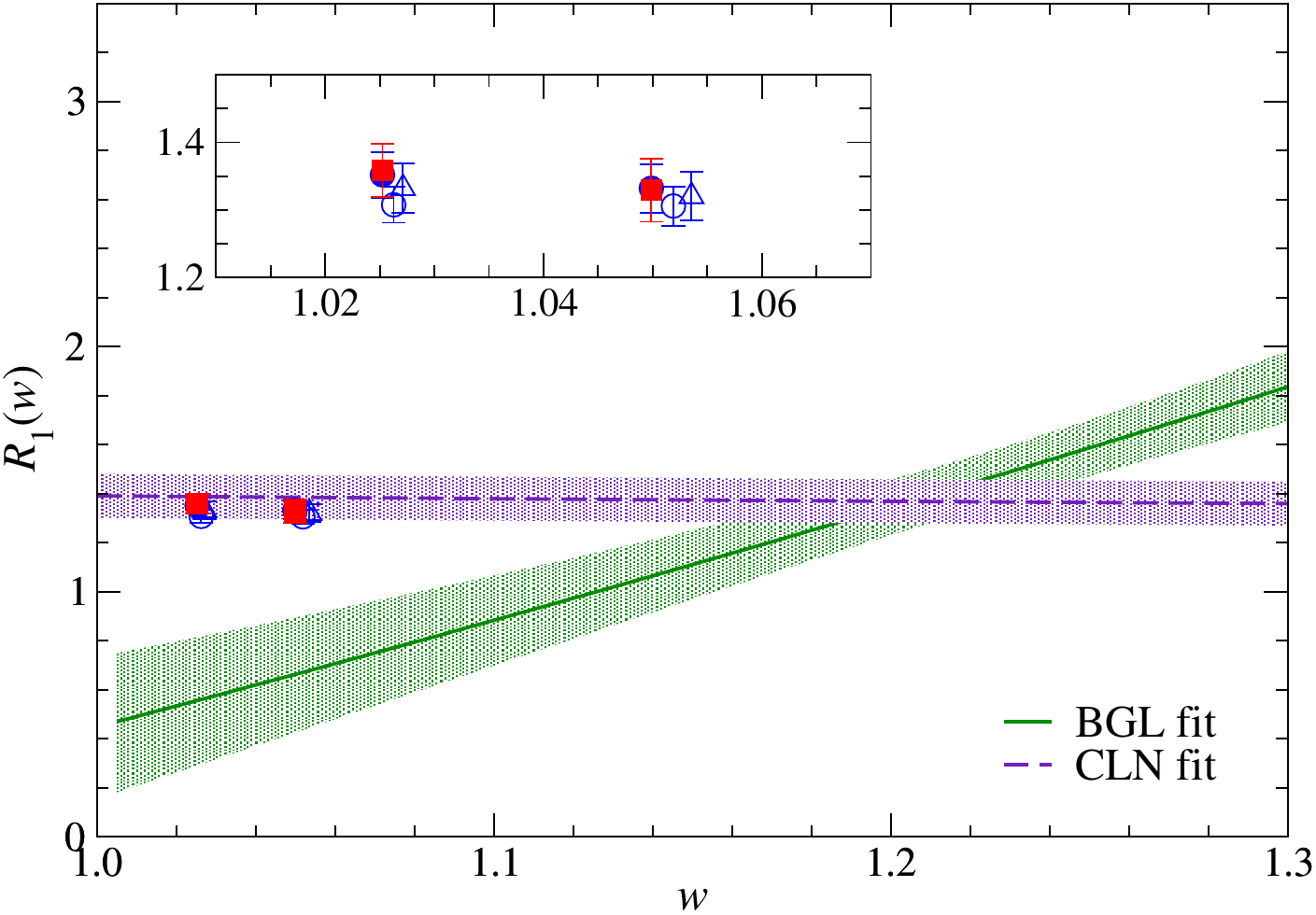}
  \caption{
    $R_1(w)\equiv h_V(w)/h_{A_1}(w)$ extracted from the BGL and CLN
    fits \cite{Bernlochner:2017xyx} of the Belle data (bands).
    The lattice results by Kaneko {\it et al.} \cite{Kaneko:2018mcr}
    are shown by the
    points (and also magnified in the inset).
    A plot from \cite{Kaneko:2018mcr}.
    \vspace*{-2mm}
  }
  \label{fig:r1_vs_w}
\end{figure}

The new observation by 
\cite{Bigi:2017njr,Grinstein:2017nlq,Bernlochner:2017xyx}
is that the BGL fit yields a higher value of $|V_{cb}|$, which is
consistent with the inclusive determination.
Since the BGL ansatz is more general, and thus more model-independent,
one might argue that this is the solution of the $|V_{cb}|$ puzzle.
The situation is, however, not that simple.
The same BGL fit also gives results on the parameters for which an
estimate from HQET at $O(1/m)$ is available,
and the results indicate unnaturally large $O(1/m)$ corrections.
An example is shown in Figure~\ref{fig:r1_vs_w}, where 
$R_1(w)\equiv h_V(w)/h_{A_1}(w)$
is plotted.
($h_V(w)$ and $h_{A_1}(w)$ are form factors to represent some
kinematical structures of $B\to D^*\ell\nu$.)
This ratio becomes unity in the heavy quark limit, and the leading
correction is expected to be of $O(\bar{\Lambda}/m_c)$,
while the plot shows that
the BGL fit of the Belle data \cite{Bernlochner:2017xyx} leads to an 
$O(1)$ deviation from 1.

More extensive test of the heavy quark symmetry relations among form
factors is necessary on both experiment and lattice, which was also
emphasized by \cite{Bigi:2017jbd}.
One example is $R_1(w)$ shown in Figure~\ref{fig:r1_vs_w}, where the
results of lattice calculation by Kaneko {\it et al.} 
\cite{Kaneko:2018mcr} are plotted
together with the BGL and CLN fits of \cite{Bernlochner:2017xyx}.
It is clear that the lattice results prefer the CLN fit,
and the latest BGL fits that yield the higher value of $|V_{cb}|$ 
are not totally satisfactory.
I should emphasize that this test is only about one parameter out of
several parameters to describe the relevant form factors,
and one cannot say from this result alone that the CLN fit is
prefered.

To summarize, the situation of the $|V_{cb}|$ determination is yet
unclear.
The lattice calculation will be the main theoretical source of
information to resolve the puzzle.
In order to fully understand the situation, the lattice inputs for the
form factor shape, not just the value in the zero-recoil limit,
will play a crucial role.
Several lattice projects are working toward this direction.
At this conference, the status is reported by
RBC/UKQCD (Witzel {\it et al.} \cite{Witzel_lat18}),
JLQCD (Kaneko {\it et al.} \cite{Kaneko:2018mcr}),
HPQCD (McLean {\it et al.} \cite{McLean:2019jll}),
Fermilab/MILC (Vaquero {\it et al.} \cite{Vaquero_lat18}, see also  \cite{Aviles-Casco:2017nge}),
as well as the Seoul group (Park {\it et al.} \cite{Bhattacharya:2018gan}).
Thorough consistency test among the experimental data, lattice
calculation and phenomenological estimate with heavy quark effective
theory would be important to finally settle the issue.

\section{More challenges}
\label{sec:challenges}

\subsection{Short-distance physics}
So far, most lattice calculations have focused on the properties of
ground-state hadrons, such as their masses or form factors.
In the real world, there are far more states of excited hadrons or
scattering hadrons.
In the perturbative analysis of QCD, one considers high-energy or
short-distance quark (and gluon) interactions and relates them to the
experimental data after summing over all possible hadronic final states.
The best-known example is the $R$-ratio, {\it i.e.} the ratio of
the $e^+e^-\to q\bar{q}$ cross section to $e^+e^-\to\mu^+\mu^-$.
Using the analyticity of the vacuum polarization function 
$\Pi(q^2)$, defined through
$(q_\mu q_\nu-q^2g_{\mu\nu})\Pi(q^2)=i\int d^4x\, e^{iqx}
\langle 0|T[j_\mu(x)j_\nu(0)]|0\rangle$,
one can write down the equality
\begin{equation}
  \label{eq:charmonium_sum_rule}
  \frac{1}{n!}\left(\frac{\partial}{\partial q^2}\right)^n
  \left.\Pi(q^2)\right|_{q^2=0}
  =
  \frac{1}{\pi}
  \int_{s_{\mathrm{min}}}^\infty ds 
  \frac{\mathrm{Im}\Pi(s)}{s^{n+1}}.
\end{equation}
Here, let us restrict ourselves to the $c\bar{c}$ final states.
The right-hand side contains the imaginary part $\mathrm{Im}\Pi(s)$,
which is related to the sum over all possible final states of
$e^+e^-\to c\bar{c}$.
The left-hand side is the vacuum polarization function at
$q^2=0$, far below the singularities --- poles and cuts.
Since $\Pi(q^2=0)$ is a short-distance quantity characterized by the
length scale of $1/m_c$, one can use the perturbative expansion to
evaluate it. 
This is the principle of the quarkonium sum rules, 
a version of the QCD sum rule 
\cite{Shifman:1978bx,Shifman:1978by}
applied for quarkonium.
The method has been used to determine the charm quark mass, for
instance, using the experimental data for $e^+e^-\to c\bar{c}$
(see \cite{Chetyrkin:2009fv,Dehnadi:2011gc} for the most recent works).

The same perturbative expansion of the charmonium vacuum polarization
function may be compared with the {\it lattice} data.
Namely, one can replace the experimental data by the lattice results
obtained directly at $q^2=0$.
The calculation is free from ultraviolet divergences when $n>1$.
The derivatives in terms of $q^2$ in 
(\ref{eq:charmonium_sum_rule})
are transformed to the temporal moments of the charmonium correlator
on the lattice, $\sum_t t^{2n} G(t)$,
which is straightforward to construct from the lattice data of the
correlator $G(t)$.
Since $G(t)$ decays exponentially in Euclidean time $t$ by the
charmonium mass, 
the sum is saturated by the short-distance region $t\sim n/m_c$.

The pioneering work of the HPQCD collaboration 
\cite{Allison:2008xk,McNeile:2010ji,Chakraborty:2014aca}
opened a new application of lattice QCD for short-distance
quantities.
Fundamental parameters in QCD, $\alpha_s(\mu)$ and $m_c(\mu)$, defined
in the $\overline{\mathrm{MS}}$ scheme were extracted directly from
the equality between the lattice data and perturbative expansion for $d^n\Pi(q^2)/dq^{2n}$.
(The same method has also been utilized more recently by
\cite{Nakayama:2016atf,Maezawa:2016vgv}.)

It also implies that one can compare the lattice data obtained in the
short-distance region with the experimental data, without recourse to
the perturbative expansion.
Such a test has been performed and an agreement between the lattice
results and experimental data has been confirmed
\cite{Donald:2012ga,Nakayama:2016atf}.
Thus, triangular links connecting three sectors, 
{\it i.e.}
experimental data, perturbation theory and lattice calculation,
are established for this quantity, which is yet another evidence of
the success of QCD.

The study of short-distance quantities necessarily involves a sum over
hadronic final states.
In this sense, the relevant process is {\it inclusive}.
Namely, one does not specify any particular final state but treats 
all possible final states with a given quantum number added with some
weight over a momentum variable.
Such analysis is not limited to the $R$-ratio (or equivalently the
vacuum polarization function), and here I would like to discuss an
application to inclusive semi-leptonic decays.

In order to be specific, let us consider the inclusive decay process 
$B\to X_c\ell\nu$ with $X_c$ 
representing any possible state with a charm quark, such as
$D^{(*)}$, $D\pi$, $D\pi\pi$, and so on.
This process is more complicated than $e^+e^-\to c\bar{c}$, as
there are two independent kinematical variables $q^2$ and $p_B\cdot q$.
Here, $q^\mu$ is a momentum transfer to the lepton pair $\ell\nu$
and $p_B^\mu$ denotes the momentum of the initial $B$ meson.

The partial decay rate is proportional to $|V_{cb}|^2$ as well as to
the structure function
\begin{equation}
  W_{\mu\nu}=\sum_{X_c} (2\pi)^3\delta^{(4)}(p_B-q-p_X)
  \frac{1}{2M_B}
  \langle B(p_B)|J_\mu^\dagger(0)|X_c\rangle
  \langle X_c|J_\nu(0)|B(p_B)\rangle,
  \label{eq:structure_function}
\end{equation}
where the sum is over all possible final states $X_c$ with momentum
specified by $p_X=p_B-q$ and
$J_\mu$ is the weak current $\bar{c}\gamma_\mu(1-\gamma_5)b$.
Using the optical theorem
$-\frac{1}{\pi}\mathrm{Im} T = W$,
one can relate the structure function 
(\ref{eq:structure_function})
to the forward scattering matrix element 
\begin{equation}
  T_{\mu\nu} = i\int d^4x\, e^{-iqx}
  \frac{1}{2M_B}
  \langle B(p_B)| T\{ J_\mu^\dagger(x)J_\nu(0) \}|B(p_B)\rangle.
\end{equation}
An important question is, then, whether this matrix element is
calculable on the lattice.

In the region of $q^2$ and $p_B\cdot q$ where the physical processes
occur, the matrix element $T_{\mu\nu}$ develops an imaginary part,
which is not easily accessible on the Euclidean lattice.
Instead, we may consider the region of $p_B\cdot q$ for which the energy
injected to the final $X_c$ state is not sufficient to produce real
states.
In other words, for the lattice calculation we consider the
kinematical region
$v\cdot (p_B-q)< M_D$.
(Here, $v^\mu=p_B^\mu/M_B$ denotes the four-velocity of the initial $B$
meson, so that the inner product $v\cdot (p_B-q)$ represents the
energy given to $X_c$.)
The matrix element in this unphysical kinematical region
may be related to the physical decay amplitude, {\it i.e.} the
imaginary part of $T_{\mu\nu}$, using Cauchy's integral of the form
\begin{equation}
  T(v\cdot q) 
  =
  \int_{-\infty}^{(v\cdot q)_{\mathrm{max}}}
  \frac{d(v\cdot q')}{\pi}
  \frac{\mathrm{Im} T(v\cdot q')}{v\cdot q'-v\cdot q},
  \label{eq:Cauchy}
\end{equation}
where I explicitly write the dependence on $v\cdot q$ while assuming
a fixed $q^2$.

Lattice calculation of the relevant matrix element is straightforward
though more costly than the standard form factor calculations.
One needs to calculate four-point functions to obtain
\begin{equation}
  C_{\mu\nu}^{JJ}(t;\mathbf{q}) =
  \int d^3\mathbf{x}\, e^{i\mathbf{q}\cdot\mathbf{x}}
  \frac{1}{2M_B}
  \langle B(\mathbf{0})| 
  J_\mu^\dagger(\mathbf{x},t) J_\nu(0)
  |B(\mathbf{0})\rangle.
\end{equation}
It is a function of a time-separation between the two inserted
currents $J_\mu^\dagger$ and $J_\nu$.
Then, the matrix element at the unphysical kinematical point may be
constructed using a ``Fourier'' (or Laplace) transform
\begin{equation}
  T_{\mu\nu}^{JJ}(\omega,\mathbf{q}) =
  \int_0^\infty dt \, e^{\omega t} C_{\mu\nu}^{JJ}(t;\mathbf{q}).
\end{equation}
Then, it may be compared with the physical amplitude through the
relation (\ref{eq:Cauchy}).
This is the strategy proposed in \cite{Hashimoto:2017wqo}.
(The same strategy may also be applied to the study of (not-so) deep
inelastic scattering. See also \cite{Chambers:2017dov}.)

\begin{figure}[tbp]
  \centering
  \vspace*{-7mm}
  \includegraphics[width=17cm]{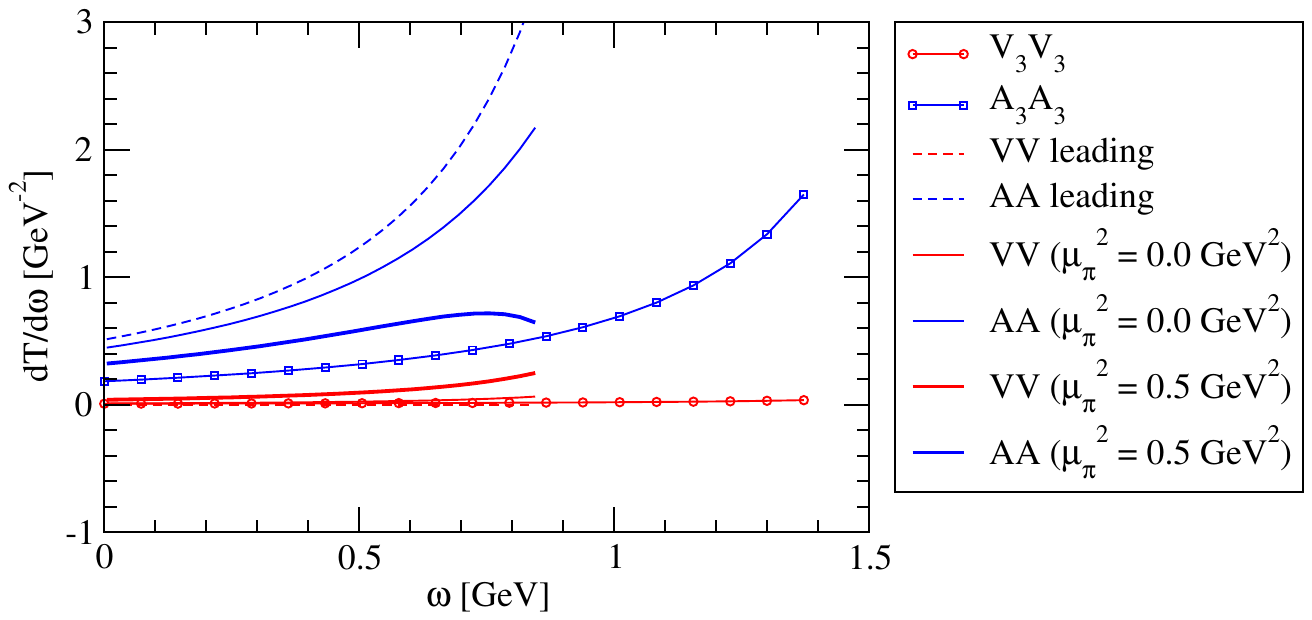}
  \vspace*{-50mm}
  \caption{
    Forward-scattering matrix elements $dT_{kk}/d\omega$ as a function
    of the energy $\omega$ injected to the final charmed state.
    Lattice data are plotted for vector $V_k$ (red circles) and
    axial-vector $A_k$ (blue squares) channels.
    Spatial recoil momentum is set to zero.
    Curves are expectations from heavy quark expansion.
    See the text for details.
  }
  \label{fig:incl}
\end{figure}

Results for $dT_{kk}/d\omega$ are plotted in Figure~\ref{fig:incl}.
Lattice data for the spatial vector (red circles) and axial-vector
(blue squares) channels are plotted together with the expectation from 
the heavy quark expansion to $O(1/m^2)$
\cite{Manohar:1993qn,Blok:1993va}.
The $b$ quark mass is lower than its physical value,
$m_b\sim 2.4 m_c$, 
and we set the spatial recoil momentum $\mathbf{q}$ to zero. 
The axial-vector channel shows a significant contribution from
the $D^*$ intermediate state.
Indeed, it develops a pole $\sim 1/(\omega-m_{D^*})^2$ towards larger
$\omega$.
The vector channel, on the hand, is nearly vanishing. 
(Precisely speaking, the vector channel is small but non-zero
representing the excited-state $D^{**}$ contribution.)

The lattice results are compared with the expectation from the heavy
quark expansion \cite{Manohar:1993qn,Blok:1993va}
(curves in Figure~\ref{fig:incl}).
The dashed curves are those of the leading order, while the solid
curves includes the corrections of $O(1/m_b^2)$.
At this order, two parameters characterizing the $B$ meson bound state
appear, {\it i.e.}
$\mu_\pi^2 = \frac{1}{2M_B}
\langle B|\bar{b}(i\vec{D})^2b|B\rangle$
and
$\mu_G^2 = \frac{1}{2M_B}
\langle B|\bar{b}\frac{i}{2}\sigma_{\mu\nu}G^{\mu\nu}b|B\rangle$.
We set $\mu_G^2$ to the value determined from the $B$-$B^*$ mass
splitting, while $\mu_\pi^2$ is more uncertain and we took two nominal
values 0~GeV$^2$ (thin lines) and 0.5~GeV$^2$ (thick lines).
As can be seen clearly in the plot, the leading order estimate is far
apart from the lattice data, and there is a trend that the $1/m^2$
correction makes them closer.
It would be interesting to see how the next order works, and, more
importantly, the perturbative corrections are to be included to have
more realistic comparison.
The work in that direction is underway \cite{Hashimoto_lat18}.

There is another proposal for the lattice study of inclusive decay.
It utilizes the lattice correlator $C_{\mu\nu}^{JJ}(t;\mathbf{q})$ in
a different way. 
If one can solve the inverse problem
\begin{equation}
  C(t) = \int_0^\infty d\omega\, e^{-\omega t}
  \left( -\frac{1}{\pi} \mathrm{Im} T(\omega) \right),
\end{equation}
the physical amplitude $\mathrm{Im} T(\omega)$ may be extracted from
the Euclidean lattice data $C(t)$.
In practice, this is not possible without having infinitely dense (and
precise) data available for $C(t)$.
Instead, one may try to extract $T(\omega)$ smeared over some small
interval. 
A systematic way of doing this was proposed by \cite{Hansen:2017mnd}.

\subsection{Long-distance effects}
Another interesting area where lattice calculation may give a
significant impact is a study of long-distance effects.
A well-known example is the $B\to K\ell^+\ell^-$ decay, which may have
a significant contribution from intermediate $b\to s\bar{c}c$ state
due to resonances $J/\psi$, $\psi'$, etc.
Since they can occur through a Cabbibo-allowed operator
${\cal O}_2=(\bar{s}\gamma_\mu P_L c)(\bar{c}\gamma^\mu P_L b)$, 
the corresponding amplitude is large especially near the resonances.
In the experimental analysis one therefore eliminate the regions of
$q^2$, invariant mass squared of the final-state leptons,
close to the corresponding resonances.
It is not entirely clear, however, how much effects are left outside
of these regions, and lattice calculation may shed new light on this
problem. 
(For a review of the situation, see, for instance
\cite{Lyon:2014hpa}.
It also emphasizes that the factorization approximation doesn't 
reproduce the available experimental data.)

Lattice study of the long-distance effects of this type has been
initiated by the RBC/UKQCD collaboration
\cite{Christ:2015aha,Christ:2016mmq}
for $K\to\pi\ell^+\ell^-$.
The problem is simpler for $B\to K\ell^+\ell^-$ as long as the effect
of the $c\bar{c}$ contributions is concerned.
Essentially, one needs to calculate the matrix element of the form
\begin{equation}
  \label{eq:4pt}
  \int_0^\infty dt\, e^{\omega t}
  \int d^3\mathbf{x}\, e^{-i\mathbf{q}\cdot\mathbf{x}}
  \langle K(\mathbf{q})|
  J_\mu^{(\mathrm{em})}(t,\mathbf{x}) {\cal H}(0)
  |B(\mathbf{0})\rangle,
\end{equation}
where the weak effective Hamiltonian ${\cal H}$ relevant to the
process of interest contains the four-fermion operator
$(\bar{c}\gamma_\mu P_L b)(\bar{s}\gamma_\mu P_L c)$
as well as
$(\bar{c}\gamma_\mu P_L c)(\bar{s}\gamma_\mu P_L b)$.
The integral over $t$ with a factor $e^{\omega t}$ specifies the
energy $\omega$ inserted to the electromagnetic current
$J_\mu^{(\mathrm{em})}=\bar{c}\gamma_\mu c$.
In order that the integral stays finite, the value of $\omega$ must be
lower than the energy of $J/\psi$.

Before analyzing the amplitude (\ref{eq:4pt}), it is interesting to
see how well the factorization assumption approximates the matrix
element 
$\langle K(\mathbf{q})|J_\mu^{(\mathrm{em})}(x) {\cal H}(0)
|B(\mathbf{0})\rangle$.
Here, the factorization assumption implies that a matrix element of
complex process approximated by a product of simpler matrix
elements, {\it e.g.}
\begin{equation}
  \langle K(\mathbf{q})|J_\mu^{(\mathrm{em})}(x) 
  (\bar{c}\gamma_\mu P_L c)(\bar{s}\gamma_\mu P_L b)(0)
  |B(\mathbf{0})\rangle
  \simeq
  \langle K(\mathbf{q})|\bar{s}\gamma_\mu P_L b(0)
  |B(\mathbf{0})\rangle
  \langle 0| J_\mu^{(\mathrm{em})}(x) \bar{c}\gamma_\mu P_L c(0)
  |0\rangle.
\end{equation}
Here the $B\to K$ form factors and charmonium decay constant are used
to express the more complicated process of
$b\to sc\bar{c}\to s\ell^+\ell^-$.
It may introduce an uncontrollable systematic error, as it ignores for
instance the effect of rescattering of the final state $K$ with the
charmonium. 

By a lattice calculation, it turned out that the factorization is well
satisfied for the operator 
${\cal O}_1=(\bar{c}\gamma_\mu P_L c)(\bar{s}\gamma_\mu P_L b)$,
while an $O(1)$ violation is observed for 
${\cal O}_2=(\bar{c}\gamma_\mu P_L b)(\bar{s}\gamma_\mu P_L c)$.
Namely, a ratio of the matrix elements of ${\cal O}_1$ and 
${\cal O}_2$ is expected to be 1/3 in the factorization approximation,
while a preliminary lattice data is more like zero
\cite{Nakayama:2019eth}.
A strong violation of the factorization approximation was found for
the $\Delta I=3/2$ amplitude of $K\to\pi\pi$ and it may be a key for
an understanding of the $\Delta I=1/2$ rule
\cite{Boyle:2012ys}.
It would therefore be interesting to see how the subtle details of
the strong interaction affects the amplitude of $B\to K\ell^+\ell^-$.

\subsection{Non-local current insertions}
There are physical quantities that are related to a matrix element of
bilocal operators.
Both of the last two applications, the forward-scattering matrix
element for inclusive decays and the non-local current insertions for 
$B\to K\ell\ell$,
are examples of such quantities.
There are more examples related to interesting physics
applications. 

One is a matrix element corresponding to the process
$B\to\ell\nu\gamma$.
This decay mode is special because by adding a photon one can avoid
the helicity suppression of the leptonic decay $B\to\ell\nu$.
It could therefore be significant even though one has to pay the
penalty of $\alpha=1/137$.
It may also provide an interesting testing ground for the lepton
flavor universality.
A lattice study of this decay mode has been presented at this conference
\cite{Soni_lat18}. 

Another such (and related) example is the calculation of the QED
correction to the leptonic and semi-leptonic processes.
The formulation to calculate the QED correction has been developed for
pion decay constant
\cite{Carrasco:2015xwa,Lubicz:2016xro,Giusti:2017dwk}.
Extension of the idea to the heavier mesons is discussed at this
conference \cite{Giusti:2018guw}.

\section{Conclusions}
Lattice calculation for heavy flavor physics reached the stage of
enabling precise calculations at a percent level.
This is an integral part of the program to search for any limitations of
the Standard Model.
Experiments, such as LHCb and Belle II, are going to produce a lot of
precise data in the coming years, and the lattice calculation has to
follow by improving its precision to maximize the power of the new
physics search.

The role of lattice calculation is not limited to improving the
precision. 
There are many physical processes which would be more useful once the
hadronic uncertainty is made under control.
The long-distance effect to $B\to K\ell^+\ell^-$ is one such example.
By extending its application, lattice calculation would play the role
to expand the horizon of flavor physics.

\section*{Acknowledgments}

I thank 
C.~Bouchard, B.~Colquhoun, C.~Davies, A.~El-Khadra,
P.~Gambino, A.~Juttner, T.~Kaneko, J.~Komijani, Z.~Ligeti,
K.~Nakayama, C.~Sachrajda, A.~Soni, A.~Vaquero 
for their inputs and discussions.
This work is supported in part by JSPS KAKENHI Grant Number 18H03710.

\bibliographystyle{JHEP-2}
\bibliography{shoji-ref}

\providecommand{\href}[2]{#2}\begingroup\raggedright\begin{thebibliography}{10}

\bibitem{Amhis:2016xyh}
{\bf HFLAV} Collaboration, Y.~Amhis {\em et.~al.}, {\it {Averages of
  $b$-hadron, $c$-hadron, and $\tau$-lepton properties as of summer 2016}},
  {\em Eur. Phys. J.} {\bf C77} (2017), no.~12 895
  [\href{http://arXiv.org/abs/1612.07233}{{\tt 1612.07233}}].

\bibitem{HFLAV_BtoDtaunu}
{\bf Heavy Flavour Averaging Group, Semileptonic B decays subgroup}
  Collaboration, C.~Bozzi, M.~Rotondo, V.~Luth, J.~Dingfelder, C.~Schwanda and
  P.~Urquijo, {\it {Average of $R(D)$ and $R(D^*)$ for Summer 2018}}, .

\bibitem{Lattice:2015rga}
{\bf MILC} Collaboration, J.~A. Bailey {\em et.~al.}, {\it {$B\to D\ell\nu$
  form factors at nonzero recoil and $|V_{cb}|$ from 2+1-flavor lattice QCD}},
  {\em Phys. Rev.} {\bf D92} (2015), no.~3 034506
  [\href{http://arXiv.org/abs/1503.07237}{{\tt 1503.07237}}].

\bibitem{Na:2015kha}
{\bf HPQCD} Collaboration, H.~Na, C.~M. Bouchard, G.~P. Lepage, C.~Monahan and
  J.~Shigemitsu, {\it {$B \rightarrow D l \nu$ form factors at nonzero recoil
  and extraction of $|V_{cb}|$}},  {\em Phys. Rev.} {\bf D92} (2015), no.~5
  054510 [\href{http://arXiv.org/abs/1505.03925}{{\tt 1505.03925}}]. [Erratum:
  Phys. Rev.D93,no.11,119906(2016)].

\bibitem{Aaij:2014ora}
{\bf LHCb} Collaboration, R.~Aaij {\em et.~al.}, {\it {Test of lepton
  universality using $B^{+}\rightarrow K^{+}\ell^{+}\ell^{-}$ decays}},  {\em
  Phys. Rev. Lett.} {\bf 113} (2014) 151601
  [\href{http://arXiv.org/abs/1406.6482}{{\tt 1406.6482}}].

\bibitem{Aaij:2017vbb}
{\bf LHCb} Collaboration, R.~Aaij {\em et.~al.}, {\it {Test of lepton
  universality with $B^{0} \rightarrow K^{*0}\ell^{+}\ell^{-}$ decays}},  {\em
  JHEP} {\bf 08} (2017) 055 [\href{http://arXiv.org/abs/1705.05802}{{\tt
  1705.05802}}].

\bibitem{Aaij:2015oid}
{\bf LHCb} Collaboration, R.~Aaij {\em et.~al.}, {\it {Angular analysis of the
  $B^{0} \to K^{*0} \mu^{+} \mu^{-}$ decay using 3 fb$^{-1}$ of integrated
  luminosity}},  {\em JHEP} {\bf 02} (2016) 104
  [\href{http://arXiv.org/abs/1512.04442}{{\tt 1512.04442}}].

\bibitem{Wehle:2016yoi}
{\bf Belle} Collaboration, S.~Wehle {\em et.~al.}, {\it
  {Lepton-Flavor-Dependent Angular Analysis of $B\to K^\ast \ell^+\ell^-$}},
  {\em Phys. Rev. Lett.} {\bf 118} (2017), no.~11 111801
  [\href{http://arXiv.org/abs/1612.05014}{{\tt 1612.05014}}].

\bibitem{Sirunyan:2017dhj}
{\bf CMS} Collaboration, A.~M. Sirunyan {\em et.~al.}, {\it {Measurement of
  angular parameters from the decay $\mathrm{B}^0 \to \mathrm{K}^{*0} \mu^+
  \mu^-$ in proton-proton collisions at $\sqrt{s} = $ 8 TeV}},  {\em Phys.
  Lett.} {\bf B781} (2018) 517--541
  [\href{http://arXiv.org/abs/1710.02846}{{\tt 1710.02846}}].

\bibitem{Aaboud:2018krd}
{\bf ATLAS} Collaboration, M.~Aaboud {\em et.~al.}, {\it {Angular analysis of
  $B^0_d \rightarrow K^{*}\mu^+\mu^-$ decays in $pp$ collisions at $\sqrt{s}=
  8$ TeV with the ATLAS detector}},
  \href{http://arXiv.org/abs/1805.04000}{{\tt 1805.04000}}.

\bibitem{Du:2015tda}
D.~Du, A.~X. El-Khadra, S.~Gottlieb, A.~S. Kronfeld, J.~Laiho, E.~Lunghi, R.~S.
  Van~de Water and R.~Zhou, {\it {Phenomenology of semileptonic B-meson decays
  with form factors from lattice QCD}},  {\em Phys. Rev.} {\bf D93} (2016),
  no.~3 034005 [\href{http://arXiv.org/abs/1510.02349}{{\tt 1510.02349}}].

\bibitem{Aoki:2016frl}
S.~Aoki {\em et.~al.}, {\it {Review of lattice results concerning low-energy
  particle physics}},  {\em Eur. Phys. J.} {\bf C77} (2017), no.~2 112
  [\href{http://arXiv.org/abs/1607.00299}{{\tt 1607.00299}}].

\bibitem{Aoki:2019cca}
S.~Aoki {\em et.~al.}, {\it {FLAG Review 2019}},
  \href{http://arXiv.org/abs/1902.08191}{{\tt 1902.08191}}.

\bibitem{Bazavov:2017lyh}
A.~Bazavov {\em et.~al.}, {\it {$B$- and $D$-meson leptonic decay constants
  from four-flavor lattice QCD}},  {\em Phys. Rev.} {\bf D98} (2018), no.~7
  074512 [\href{http://arXiv.org/abs/1712.09262}{{\tt 1712.09262}}].

\bibitem{Bazavov:2011aa}
{\bf Fermilab Lattice, MILC} Collaboration, A.~Bazavov {\em et.~al.}, {\it {B-
  and D-meson decay constants from three-flavor lattice QCD}},  {\em Phys.
  Rev.} {\bf D85} (2012) 114506 [\href{http://arXiv.org/abs/1112.3051}{{\tt
  1112.3051}}].

\bibitem{Christ:2014uea}
N.~H. Christ, J.~M. Flynn, T.~Izubuchi, T.~Kawanai, C.~Lehner, A.~Soni, R.~S.
  Van~de Water and O.~Witzel, {\it {B-meson decay constants from 2+1-flavor
  lattice QCD with domain-wall light quarks and relativistic heavy quarks}},
  {\em Phys. Rev.} {\bf D91} (2015), no.~5 054502
  [\href{http://arXiv.org/abs/1404.4670}{{\tt 1404.4670}}].

\bibitem{Aoki:2014nga}
Y.~Aoki, T.~Ishikawa, T.~Izubuchi, C.~Lehner and A.~Soni, {\it {Neutral $B$
  meson mixings and $B$ meson decay constants with static heavy and domain-wall
  light quarks}},  {\em Phys. Rev.} {\bf D91} (2015), no.~11 114505
  [\href{http://arXiv.org/abs/1406.6192}{{\tt 1406.6192}}].

\bibitem{Na:2012kp}
H.~Na, C.~J. Monahan, C.~T.~H. Davies, R.~Horgan, G.~P. Lepage and
  J.~Shigemitsu, {\it {The $B$ and $B_s$ Meson Decay Constants from Lattice
  QCD}},  {\em Phys. Rev.} {\bf D86} (2012) 034506
  [\href{http://arXiv.org/abs/1202.4914}{{\tt 1202.4914}}].

\bibitem{Dowdall:2013tga}
{\bf HPQCD} Collaboration, R.~J. Dowdall, C.~T.~H. Davies, R.~R. Horgan, C.~J.
  Monahan and J.~Shigemitsu, {\it {B-Meson Decay Constants from Improved
  Lattice Nonrelativistic QCD with Physical u, d, s, and c Quarks}},  {\em
  Phys. Rev. Lett.} {\bf 110} (2013), no.~22 222003
  [\href{http://arXiv.org/abs/1302.2644}{{\tt 1302.2644}}].

\bibitem{Follana:2006rc}
{\bf HPQCD, UKQCD} Collaboration, E.~Follana, Q.~Mason, C.~Davies,
  K.~Hornbostel, G.~P. Lepage, J.~Shigemitsu, H.~Trottier and K.~Wong, {\it
  {Highly improved staggered quarks on the lattice, with applications to charm
  physics}},  {\em Phys. Rev.} {\bf D75} (2007) 054502
  [\href{http://arXiv.org/abs/hep-lat/0610092}{{\tt hep-lat/0610092}}].

\bibitem{ElKhadra:1996mp}
A.~X. El-Khadra, A.~S. Kronfeld and P.~B. Mackenzie, {\it {Massive fermions in
  lattice gauge theory}},  {\em Phys. Rev.} {\bf D55} (1997) 3933--3957
  [\href{http://arXiv.org/abs/hep-lat/9604004}{{\tt hep-lat/9604004}}].

\bibitem{McNeile:2011ng}
C.~McNeile, C.~T.~H. Davies, E.~Follana, K.~Hornbostel and G.~P. Lepage, {\it
  {High-Precision $f_{B_s}$ and HQET from Relativistic Lattice QCD}},  {\em
  Phys. Rev.} {\bf D85} (2012) 031503
  [\href{http://arXiv.org/abs/1110.4510}{{\tt 1110.4510}}].

\bibitem{Lepage:1992tx}
G.~P. Lepage, L.~Magnea, C.~Nakhleh, U.~Magnea and K.~Hornbostel, {\it
  {Improved nonrelativistic QCD for heavy quark physics}},  {\em Phys. Rev.}
  {\bf D46} (1992) 4052--4067 [\href{http://arXiv.org/abs/hep-lat/9205007}{{\tt
  hep-lat/9205007}}].

\bibitem{Heitger:2003nj}
{\bf ALPHA} Collaboration, J.~Heitger and R.~Sommer, {\it {Nonperturbative
  heavy quark effective theory}},  {\em JHEP} {\bf 02} (2004) 022
  [\href{http://arXiv.org/abs/hep-lat/0310035}{{\tt hep-lat/0310035}}].

\bibitem{Blossier:2010jk}
B.~Blossier, M.~della Morte, N.~Garron and R.~Sommer, {\it {HQET at order
  $1/m$: I. Non-perturbative parameters in the quenched approximation}},  {\em
  JHEP} {\bf 06} (2010) 002 [\href{http://arXiv.org/abs/1001.4783}{{\tt
  1001.4783}}].

\bibitem{Blossier:2010vz}
{\bf Alpha} Collaboration, B.~Blossier, M.~Della~Morte, N.~Garron, G.~von
  Hippel, T.~Mendes, H.~Simma and R.~Sommer, {\it {HQET at order $1/m$: II.
  Spectroscopy in the quenched approximation}},  {\em JHEP} {\bf 05} (2010) 074
  [\href{http://arXiv.org/abs/1004.2661}{{\tt 1004.2661}}].

\bibitem{Blossier:2010mk}
{\bf ALPHA} Collaboration, B.~Blossier, M.~Della~Morte, N.~Garron, G.~von
  Hippel, T.~Mendes, H.~Simma and R.~Sommer, {\it {HQET at order 1/m: III.
  Decay constants in the quenched approximation}},  {\em JHEP} {\bf 12} (2010)
  039 [\href{http://arXiv.org/abs/1006.5816}{{\tt 1006.5816}}].

\bibitem{Lepage:1989hd}
G.~P. Lepage, {\it {The Analysis of Algorithms for Lattice Field Theory}},  in
  {\em {Boulder ASI 1989:97-120}}, pp.~97--120, 1989.

\bibitem{Hashimoto:1994nd}
S.~Hashimoto, {\it {Computation of the heavy - light decay constant using
  nonrelativistic lattice QCD}},  {\em Phys. Rev.} {\bf D50} (1994) 4639--4648
  [\href{http://arXiv.org/abs/hep-lat/9403028}{{\tt hep-lat/9403028}}].

\bibitem{Rajan:2017lxk}
R.~Gupta, Y.-C. Jang, H.-W. Lin, B.~Yoon and T.~Bhattacharya, {\it {Axial
  Vector Form Factors of the Nucleon from Lattice QCD}},  {\em Phys. Rev.} {\bf
  D96} (2017), no.~11 114503 [\href{http://arXiv.org/abs/1705.06834}{{\tt
  1705.06834}}].

\bibitem{Bar:2017gqh}
O.~Bar, {\it {Multi-hadron-state contamination in nucleon observables from
  chiral perturbation theory}},  {\em EPJ Web Conf.} {\bf 175} (2018) 01007
  [\href{http://arXiv.org/abs/1708.00380}{{\tt 1708.00380}}].

\bibitem{Bar:2016uoj}
O.~Bär, {\it {Nucleon-pion-state contribution in lattice calculations of the
  nucleon charges $g_A,g_T$ and $g_S$}},  {\em Phys. Rev.} {\bf D94} (2016),
  no.~5 054505 [\href{http://arXiv.org/abs/1606.09385}{{\tt 1606.09385}}].

\bibitem{Bar:2018wco}
O.~Bär, {\it {Three-particle $N\pi\pi$ state contribution to the nucleon
  two-point function in lattice QCD}},  {\em Phys. Rev.} {\bf D97} (2018),
  no.~9 094507 [\href{http://arXiv.org/abs/1802.10442}{{\tt 1802.10442}}].

\bibitem{Bar:2018akl}
O.~Bar, {\it {Nucleon-pion-state contamination in lattice calculations of the
  axial form factors of the nucleon}},  {\em PoS} {\bf LATTICE2018} (2018) 061
  [\href{http://arXiv.org/abs/1808.08738}{{\tt 1808.08738}}].

\bibitem{Detmold:2015aaa}
W.~Detmold, C.~Lehner and S.~Meinel, {\it {$\Lambda_b \to p \ell^-
  \bar{\nu}_\ell$ and $\Lambda_b \to \Lambda_c \ell^- \bar{\nu}_\ell$ form
  factors from lattice QCD with relativistic heavy quarks}},  {\em Phys. Rev.}
  {\bf D92} (2015), no.~3 034503 [\href{http://arXiv.org/abs/1503.01421}{{\tt
  1503.01421}}].

\bibitem{Flynn:2015mha}
J.~M. Flynn, T.~Izubuchi, T.~Kawanai, C.~Lehner, A.~Soni, R.~S. Van~de Water
  and O.~Witzel, {\it {$B \to \pi \ell \nu$ and $B_s \to K \ell \nu$ form
  factors and $|V_{ub}|$ from 2+1-flavor lattice QCD with domain-wall light
  quarks and relativistic heavy quarks}},  {\em Phys. Rev.} {\bf D91} (2015),
  no.~7 074510 [\href{http://arXiv.org/abs/1501.05373}{{\tt 1501.05373}}].

\bibitem{Lattice:2015tia}
{\bf Fermilab Lattice, MILC} Collaboration, J.~A. Bailey {\em et.~al.}, {\it
  {$|V_{ub}|$ from $B\to\pi\ell\nu$ decays and (2+1)-flavor lattice QCD}},
  {\em Phys. Rev.} {\bf D92} (2015), no.~1 014024
  [\href{http://arXiv.org/abs/1503.07839}{{\tt 1503.07839}}].

\bibitem{Bahr:2017ivh}
F.~Bahr, D.~Banerjee, M.~Koren, H.~Simma and R.~Sommer, {\it {Extraction of the
  bare form factors for the semi-leptonic $B_s$ decays}},  {\em PoS} {\bf
  LATTICE2016} (2016) 295 [\href{http://arXiv.org/abs/1701.03299}{{\tt
  1701.03299}}].

\bibitem{Gelzer_lat18}
Z.~Gelzer, {\it {Semileptonic decays of $B_s$ mesons to light pseudoscalar
  mesons on four-flavor HISQ ensembles}},  {\em PoS} {\bf LATTICE2018} (2018)
  289.

\bibitem{Gelzer:2017edb}
Z.~Gelzer {\em et.~al.}, {\it {Semileptonic $B$-meson decays to light
  pseudoscalar mesons on the HISQ ensembles}},  {\em EPJ Web Conf.} {\bf 175}
  (2018) 13024 [\href{http://arXiv.org/abs/1710.09442}{{\tt 1710.09442}}].

\bibitem{Lattice:2017vqf}
{\bf Fermilab Lattice, MILC} Collaboration, Y.~Liu {\em et.~al.}, {\it {$B_s
  \to K \ell\nu$ Form Factors with 2+1 Flavors}},  {\em EPJ Web Conf.} {\bf
  175} (2018) 13008 [\href{http://arXiv.org/abs/1711.08085}{{\tt 1711.08085}}].

\bibitem{Witzel_lat18}
O.~Witzel, {\it {Semi-leptonic form factors for $B_s\to K\ell\nu$ and $B_s\to
  D_s\ell\nu$}},  {\em PoS} {\bf LATTICE2018} (2018) 290.

\bibitem{Colquhoun:2018kwj}
B.~Colquhoun, S.~Hashimoto and T.~Kaneko, {\it {Heavy quark scaling of
  $B\to\pi\ell\nu$ form factors with M\"{o}bius domain wall fermions}},  {\em
  PoS} {\bf LATTICE2018} (2018) 274
  [\href{http://arXiv.org/abs/1811.00227}{{\tt 1811.00227}}].

\bibitem{Colquhoun:2017gfi}
B.~Colquhoun, S.~Hashimoto and T.~Kaneko, {\it {$B \rightarrow \pi \ell \nu$
  with Möbius Domain Wall Fermions}},  {\em EPJ Web Conf.} {\bf 175} (2018)
  13004 [\href{http://arXiv.org/abs/1710.07094}{{\tt 1710.07094}}].

\bibitem{Bouchard_lat18}
C.~Bouchard, {\it {$B\to\pi\ell\nu$ and $B\to\pi\ell\ell$ decays with
  HISQ/NRQCD valence quarks on $N_f$ = 2+1 asqtad ensembles}},  {\em PoS} {\bf
  LATTICE2018} (2018) 288.

\bibitem{Monahan:2018lzv}
C.~J. Monahan, C.~M. Bouchard, G.~P. Lepage, H.~Na and J.~Shigemitsu, {\it
  {Form factor ratios for $B_s \rightarrow K \, \ell \, \nu$ and $B_s
  \rightarrow D_s \, \ell \, \nu$ semileptonic decays and $|V_{ub}/V_{cb}|$}},
  \href{http://arXiv.org/abs/1808.09285}{{\tt 1808.09285}}.

\bibitem{Lubicz:2018scy}
V.~Lubicz, L.~Riggio, G.~Salerno and S.~Simula, {\it {Hypercubic effects in
  semileptonic decays of heavy mesons, toward $B \to \pi \ell \nu$, with
  $N_f=2+1+1$ Twisted fermions}},  {\em PoS} {\bf LATTICE2018} (2018) 287
  [\href{http://arXiv.org/abs/1811.10268}{{\tt 1811.10268}}].

\bibitem{Lubicz:2018rfs}
{\bf ETM} Collaboration, V.~Lubicz, L.~Riggio, G.~Salerno, S.~Simula and
  C.~Tarantino, {\it {Tensor form factor of $D \to \pi(K) \ell \nu$ and $D \to
  \pi(K) \ell \ell$ decays with $N_f=2+1+1$ twisted-mass fermions}},  {\em
  Phys. Rev.} {\bf D98} (2018), no.~1 014516
  [\href{http://arXiv.org/abs/1803.04807}{{\tt 1803.04807}}].

\bibitem{Isgur:1989vq}
N.~Isgur and M.~B. Wise, {\it {Weak Decays of Heavy Mesons in the Static Quark
  Approximation}},  {\em Phys. Lett.} {\bf B232} (1989) 113--117.

\bibitem{Isgur:1989ed}
N.~Isgur and M.~B. Wise, {\it {WEAK TRANSITION FORM-FACTORS BETWEEN HEAVY
  MESONS}},  {\em Phys. Lett.} {\bf B237} (1990) 527--530.

\bibitem{Luke:1990eg}
M.~E. Luke, {\it {Effects of subleading operators in the heavy quark effective
  theory}},  {\em Phys. Lett.} {\bf B252} (1990) 447--455.

\bibitem{Hashimoto:1999yp}
S.~Hashimoto, A.~X. El-Khadra, A.~S. Kronfeld, P.~B. Mackenzie, S.~M. Ryan and
  J.~N. Simone, {\it {Lattice QCD calculation of $\bar{B}\to D\ell\bar{\nu}$
  decay form-factors at zero recoil}},  {\em Phys. Rev.} {\bf D61} (1999)
  014502 [\href{http://arXiv.org/abs/hep-ph/9906376}{{\tt hep-ph/9906376}}].

\bibitem{Hashimoto:2001nb}
S.~Hashimoto, A.~S. Kronfeld, P.~B. Mackenzie, S.~M. Ryan and J.~N. Simone,
  {\it {Lattice calculation of the zero recoil form-factor of $\bar{B}\to
  D^*\ell\nu$: Toward a model independent determination of $|V_{cb}|$}},  {\em
  Phys. Rev.} {\bf D66} (2002) 014503
  [\href{http://arXiv.org/abs/hep-ph/0110253}{{\tt hep-ph/0110253}}].

\bibitem{Bailey:2014tva}
{\bf Fermilab Lattice, MILC} Collaboration, J.~A. Bailey {\em et.~al.}, {\it
  {Update of $|V_{cb}|$ from the $\bar{B}\to D^*\ell\bar{\nu}$ form factor at
  zero recoil with three-flavor lattice QCD}},  {\em Phys. Rev.} {\bf D89}
  (2014), no.~11 114504 [\href{http://arXiv.org/abs/1403.0635}{{\tt
  1403.0635}}].

\bibitem{Harrison:2017fmw}
{\bf HPQCD} Collaboration, J.~Harrison, C.~Davies and M.~Wingate, {\it {Lattice
  QCD calculation of the ${{B}_{(s)}\to D_{(s)}^{*}\ell{\nu}}$ form factors at
  zero recoil and implications for ${|V_{cb}|}$}},  {\em Phys. Rev.} {\bf D97}
  (2018), no.~5 054502 [\href{http://arXiv.org/abs/1711.11013}{{\tt
  1711.11013}}].

\bibitem{Abdesselam:2017kjf}
{\bf Belle} Collaboration, A.~Abdesselam {\em et.~al.}, {\it {Precise
  determination of the CKM matrix element $\left| V_{cb}\right|$ with $\bar B^0
  \to D^{*\,+} \, \ell^- \, \bar \nu_\ell$ decays with hadronic tagging at
  Belle}},  \href{http://arXiv.org/abs/1702.01521}{{\tt 1702.01521}}.

\bibitem{Bigi:2017njr}
D.~Bigi, P.~Gambino and S.~Schacht, {\it {A fresh look at the determination of
  $|V_{cb}|$ from $B\to D^{*} \ell \nu$}},  {\em Phys. Lett.} {\bf B769} (2017)
  441--445 [\href{http://arXiv.org/abs/1703.06124}{{\tt 1703.06124}}].

\bibitem{Grinstein:2017nlq}
B.~Grinstein and A.~Kobach, {\it {Model-Independent Extraction of $|V_{cb}|$
  from $\bar{B}\rightarrow D^* \ell \overline{\nu}$}},  {\em Phys. Lett.} {\bf
  B771} (2017) 359--364 [\href{http://arXiv.org/abs/1703.08170}{{\tt
  1703.08170}}].

\bibitem{Bernlochner:2017xyx}
F.~U. Bernlochner, Z.~Ligeti, M.~Papucci and D.~J. Robinson, {\it {Tensions and
  correlations in $|V_{cb}|$ determinations}},  {\em Phys. Rev.} {\bf D96}
  (2017), no.~9 091503 [\href{http://arXiv.org/abs/1708.07134}{{\tt
  1708.07134}}].

\bibitem{Caprini:1997mu}
I.~Caprini, L.~Lellouch and M.~Neubert, {\it {Dispersive bounds on the shape of
  $\bar{B}\to D^{(*)}\ell\bar{\nu}$ form-factors}},  {\em Nucl. Phys.} {\bf
  B530} (1998) 153--181 [\href{http://arXiv.org/abs/hep-ph/9712417}{{\tt
  hep-ph/9712417}}].

\bibitem{Boyd:1995cf}
C.~G. Boyd, B.~Grinstein and R.~F. Lebed, {\it {Model independent extraction of
  $|V_{cb}|$ using dispersion relations}},  {\em Phys. Lett.} {\bf B353} (1995)
  306--312 [\href{http://arXiv.org/abs/hep-ph/9504235}{{\tt hep-ph/9504235}}].

\bibitem{Okubo:1972ih}
S.~Okubo and I.-F. Shih, {\it {Exact inequality and test of chiral sw(3) theory
  in k-l-3 decay problem}},  {\em Phys. Rev.} {\bf D4} (1971) 2020--2029.

\bibitem{Bourrely:1980gp}
C.~Bourrely, B.~Machet and E.~de~Rafael, {\it {Semileptonic Decays of
  Pseudoscalar Particles ($M \to M^\prime \ell \nu_\ell$) and Short Distance
  Behavior of Quantum Chromodynamics}},  {\em Nucl. Phys.} {\bf B189} (1981)
  157--181.

\bibitem{Kaneko:2018mcr}
{\bf JLQCD} Collaboration, T.~Kaneko, Y.~Aoki, B.~Colquhoun, H.~Fukaya and
  S.~Hashimoto, {\it {$B \to D^{(*)}\ell\nu$ form factors from $N_f\!=\!2+1$
  QCD with M\"obius domain-wall quarks}},  {\em PoS} {\bf LATTICE2018} (2018)
  311 [\href{http://arXiv.org/abs/1811.00794}{{\tt 1811.00794}}].

\bibitem{Bigi:2017jbd}
D.~Bigi, P.~Gambino and S.~Schacht, {\it {$R(D^*)$, $|V_{cb}|$, and the Heavy
  Quark Symmetry relations between form factors}},  {\em JHEP} {\bf 11} (2017)
  061 [\href{http://arXiv.org/abs/1707.09509}{{\tt 1707.09509}}].

\bibitem{McLean:2019jll}
E.~McLean, C.~T.~H. Davies, A.~T. Lytle and J.~Koponen, {\it {$B_s\to
  D_s^{(*)}l\nu$ Form Factors with Heavy HISQ Quarks}},  {\em PoS} {\bf
  LATTICE2018} (2018) 282 [\href{http://arXiv.org/abs/1901.04979}{{\tt
  1901.04979}}].

\bibitem{Vaquero_lat18}
A.~Vaquero, {\it {$B\rightarrow D^*\ell\nu$ at non-zero recoil}},  {\em PoS}
  {\bf LATTICE2018} (2018) 282.

\bibitem{Aviles-Casco:2017nge}
A.~Vaquero Avilés-Casco, C.~DeTar, D.~Du, A.~El-Khadra, A.~S. Kronfeld,
  J.~Laiho and R.~S. Van~de Water, {\it {$\overline{B}\rightarrow
  D^\ast\ell\overline{\nu}$ at Non-Zero Recoil}},  {\em EPJ Web Conf.} {\bf
  175} (2018) 13003 [\href{http://arXiv.org/abs/1710.09817}{{\tt 1710.09817}}].

\bibitem{Bhattacharya:2018gan}
T.~Bhattacharya, R.~Gupta, S.~Park, Y.~Jang, J.~Bailey, B.~Choi, H.~Jeong,
  S.~Jwa, S.~Lee, W.~Lee, J.~Pak and J.~Leem, {\it {Update on $B\to D^*\ell\nu$
  form factor at zero-recoil using the Oktay-Kronfeld action}},  {\em PoS} {\bf
  LATTICE2018} (2018) 283 [\href{http://arXiv.org/abs/1812.07675}{{\tt
  1812.07675}}].

\bibitem{Shifman:1978bx}
M.~A. Shifman, A.~I. Vainshtein and V.~I. Zakharov, {\it {QCD and Resonance
  Physics. Theoretical Foundations}},  {\em Nucl. Phys.} {\bf B147} (1979)
  385--447.

\bibitem{Shifman:1978by}
M.~A. Shifman, A.~I. Vainshtein and V.~I. Zakharov, {\it {QCD and Resonance
  Physics: Applications}},  {\em Nucl. Phys.} {\bf B147} (1979) 448--518.

\bibitem{Chetyrkin:2009fv}
K.~G. Chetyrkin, J.~H. Kuhn, A.~Maier, P.~Maierhofer, P.~Marquard,
  M.~Steinhauser and C.~Sturm, {\it {Charm and Bottom Quark Masses: An
  Update}},  {\em Phys. Rev.} {\bf D80} (2009) 074010
  [\href{http://arXiv.org/abs/0907.2110}{{\tt 0907.2110}}].

\bibitem{Dehnadi:2011gc}
B.~Dehnadi, A.~H. Hoang, V.~Mateu and S.~M. Zebarjad, {\it {Charm Mass
  Determination from QCD Charmonium Sum Rules at Order $\alpha_{s}^{3}$}},
  {\em JHEP} {\bf 09} (2013) 103 [\href{http://arXiv.org/abs/1102.2264}{{\tt
  1102.2264}}].

\bibitem{Allison:2008xk}
{\bf HPQCD} Collaboration, I.~Allison {\em et.~al.}, {\it {High-Precision
  Charm-Quark Mass from Current-Current Correlators in Lattice and Continuum
  QCD}},  {\em Phys. Rev.} {\bf D78} (2008) 054513
  [\href{http://arXiv.org/abs/0805.2999}{{\tt 0805.2999}}].

\bibitem{McNeile:2010ji}
C.~McNeile, C.~T.~H. Davies, E.~Follana, K.~Hornbostel and G.~P. Lepage, {\it
  {High-Precision c and b Masses, and QCD Coupling from Current-Current
  Correlators in Lattice and Continuum QCD}},  {\em Phys. Rev.} {\bf D82}
  (2010) 034512 [\href{http://arXiv.org/abs/1004.4285}{{\tt 1004.4285}}].

\bibitem{Chakraborty:2014aca}
B.~Chakraborty, C.~T.~H. Davies, B.~Galloway, P.~Knecht, J.~Koponen, G.~C.
  Donald, R.~J. Dowdall, G.~P. Lepage and C.~McNeile, {\it {High-precision
  quark masses and QCD coupling from $n_f=4$ lattice QCD}},  {\em Phys. Rev.}
  {\bf D91} (2015), no.~5 054508 [\href{http://arXiv.org/abs/1408.4169}{{\tt
  1408.4169}}].

\bibitem{Nakayama:2016atf}
K.~Nakayama, B.~Fahy and S.~Hashimoto, {\it {Short-distance charmonium
  correlator on the lattice with Möbius domain-wall fermion and a
  determination of charm quark mass}},  {\em Phys. Rev.} {\bf D94} (2016),
  no.~5 054507 [\href{http://arXiv.org/abs/1606.01002}{{\tt 1606.01002}}].

\bibitem{Maezawa:2016vgv}
Y.~Maezawa and P.~Petreczky, {\it {Quark masses and strong coupling constant in
  2+1 flavor QCD}},  {\em Phys. Rev.} {\bf D94} (2016), no.~3 034507
  [\href{http://arXiv.org/abs/1606.08798}{{\tt 1606.08798}}].

\bibitem{Donald:2012ga}
G.~C. Donald, C.~T.~H. Davies, R.~J. Dowdall, E.~Follana, K.~Hornbostel,
  J.~Koponen, G.~P. Lepage and C.~McNeile, {\it {Precision tests of the
  $J/{\psi}$ from full lattice QCD: mass, leptonic width and radiative decay
  rate to ${\eta}_c$}},  {\em Phys. Rev.} {\bf D86} (2012) 094501
  [\href{http://arXiv.org/abs/1208.2855}{{\tt 1208.2855}}].

\bibitem{Hashimoto:2017wqo}
S.~Hashimoto, {\it {Inclusive semi-leptonic B meson decay structure functions
  from lattice QCD}},  {\em PTEP} {\bf 2017} (2017), no.~5 053B03
  [\href{http://arXiv.org/abs/1703.01881}{{\tt 1703.01881}}].

\bibitem{Chambers:2017dov}
A.~J. Chambers, R.~Horsley, Y.~Nakamura, H.~Perlt, P.~E.~L. Rakow,
  G.~Schierholz, A.~Schiller, K.~Somfleth, R.~D. Young and J.~M. Zanotti, {\it
  {Nucleon Structure Functions from Operator Product Expansion on the
  Lattice}},  {\em Phys. Rev. Lett.} {\bf 118} (2017), no.~24 242001
  [\href{http://arXiv.org/abs/1703.01153}{{\tt 1703.01153}}].

\bibitem{Manohar:1993qn}
A.~V. Manohar and M.~B. Wise, {\it {Inclusive semileptonic B and polarized
  Lambda(b) decays from QCD}},  {\em Phys. Rev.} {\bf D49} (1994) 1310--1329
  [\href{http://arXiv.org/abs/hep-ph/9308246}{{\tt hep-ph/9308246}}].

\bibitem{Blok:1993va}
B.~Blok, L.~Koyrakh, M.~A. Shifman and A.~I. Vainshtein, {\it {Differential
  distributions in semileptonic decays of the heavy flavors in QCD}},  {\em
  Phys. Rev.} {\bf D49} (1994) 3356
  [\href{http://arXiv.org/abs/hep-ph/9307247}{{\tt hep-ph/9307247}}]. [Erratum:
  Phys. Rev.D50,3572(1994)].

\bibitem{Hashimoto_lat18}
B.~Colquhoun, P.~Gambino, S.~Hashimoto and T.~Kaneko, {\it {Inclusive decay
  structure function for $B\to X_c\ell\nu$: a comparison of a lattice
  calculation with the heavy quark expansion}},  {\em PoS} {\bf LATTICE2018}
  (2018) 307.

\bibitem{Hansen:2017mnd}
M.~T. Hansen, H.~B. Meyer and D.~Robaina, {\it {From deep inelastic scattering
  to heavy-flavor semileptonic decays: Total rates into multihadron final
  states from lattice QCD}},  {\em Phys. Rev.} {\bf D96} (2017), no.~9 094513
  [\href{http://arXiv.org/abs/1704.08993}{{\tt 1704.08993}}].

\bibitem{Lyon:2014hpa}
J.~Lyon and R.~Zwicky, {\it {Resonances gone topsy turvy - the charm of QCD or
  new physics in $b \to s \ell^+ \ell^-$?}},
  \href{http://arXiv.org/abs/1406.0566}{{\tt 1406.0566}}.

\bibitem{Christ:2015aha}
{\bf RBC, UKQCD} Collaboration, N.~H. Christ, X.~Feng, A.~Portelli and C.~T.
  Sachrajda, {\it {Prospects for a lattice computation of rare kaon decay
  amplitudes: $K\to\pi\ell^+\ell^-$ decays}},  {\em Phys. Rev.} {\bf D92}
  (2015), no.~9 094512 [\href{http://arXiv.org/abs/1507.03094}{{\tt
  1507.03094}}].

\bibitem{Christ:2016mmq}
N.~H. Christ, X.~Feng, A.~Juttner, A.~Lawson, A.~Portelli and C.~T. Sachrajda,
  {\it {First exploratory calculation of the long-distance contributions to the
  rare kaon decays $K\to\pi\ell^+\ell^-$}},  {\em Phys. Rev.} {\bf D94} (2016),
  no.~11 114516 [\href{http://arXiv.org/abs/1608.07585}{{\tt 1608.07585}}].

\bibitem{Nakayama:2019eth}
K.~Nakayama and S.~Hashimoto, {\it {Test of factorization for the long-distance
  effects from charmonium in $B\rightarrow K\ell\ell$}},  {\em PoS} {\bf
  LATTICE2018} (2018) 221 [\href{http://arXiv.org/abs/1901.08784}{{\tt
  1901.08784}}].

\bibitem{Boyle:2012ys}
{\bf RBC, UKQCD} Collaboration, P.~A. Boyle {\em et.~al.}, {\it {Emerging
  understanding of the $\Delta I = 1/2$ Rule from Lattice QCD}},  {\em Phys.
  Rev. Lett.} {\bf 110} (2013), no.~15 152001
  [\href{http://arXiv.org/abs/1212.1474}{{\tt 1212.1474}}].

\bibitem{Soni_lat18}
A.~Soni, {\it {Flavor anomalies \& the lattice}},  {\em PoS} {\bf LATTICE2018}
  (2018) 292.

\bibitem{Carrasco:2015xwa}
N.~Carrasco, V.~Lubicz, G.~Martinelli, C.~T. Sachrajda, N.~Tantalo,
  C.~Tarantino and M.~Testa, {\it {QED Corrections to Hadronic Processes in
  Lattice QCD}},  {\em Phys. Rev.} {\bf D91} (2015), no.~7 074506
  [\href{http://arXiv.org/abs/1502.00257}{{\tt 1502.00257}}].

\bibitem{Lubicz:2016xro}
V.~Lubicz, G.~Martinelli, C.~T. Sachrajda, F.~Sanfilippo, S.~Simula and
  N.~Tantalo, {\it {Finite-Volume QED Corrections to Decay Amplitudes in
  Lattice QCD}},  {\em Phys. Rev.} {\bf D95} (2017), no.~3 034504
  [\href{http://arXiv.org/abs/1611.08497}{{\tt 1611.08497}}].

\bibitem{Giusti:2017dwk}
D.~Giusti, V.~Lubicz, G.~Martinelli, C.~T. Sachrajda, F.~Sanfilippo, S.~Simula,
  N.~Tantalo and C.~Tarantino, {\it {First lattice calculation of the QED
  corrections to leptonic decay rates}},  {\em Phys. Rev. Lett.} {\bf 120}
  (2018), no.~7 072001 [\href{http://arXiv.org/abs/1711.06537}{{\tt
  1711.06537}}].

\bibitem{Giusti:2018guw}
D.~Giusti, V.~Lubicz, G.~Martinelli, C.~Sachrajda, F.~Sanfilippo, S.~Simula and
  N.~Tantalo, {\it {Radiative corrections to decay amplitudes in lattice QCD}},
   {\em PoS} {\bf LATTICE2018} (2018) 266
  [\href{http://arXiv.org/abs/1811.06364}{{\tt 1811.06364}}].

\end{thebibliography}\endgroup

\end{document}